\documentclass[twocolumn]{aastex631}

\turnoffediting

\usepackage{graphicx}	
\usepackage{amsmath}	
\usepackage{amssymb}	

\usepackage{comment}
\usepackage{hyperref}

\begin{document}
\title{Connecting the X-ray/UV variability of Fairall~9 with NICER: A Possible Warm Corona}
\author[0000-0003-1183-1574]{Ethan R. Partington}
\affiliation{Department of Physics and Astronomy, Wayne State University, 666 W.\ Hancock St, Detroit, MI, 48201, USA}

\author[0000-0002-8294-9281]{Edward M.\ Cackett}
\affiliation{Department of Physics and Astronomy, Wayne State University, 666 W.\ Hancock St, Detroit, MI, 48201, USA}

\author[0000-0001-8598-1482]{Rick Edelson} 
\affiliation{Eureka Scientific Inc., 2452 Delmer St. Suite 100, Oakland, CA 94602, USA}

\author[0000-0003-1728-0304]{Keith Horne}
\affiliation{SUPA School of Physics and Astronomy, North Haugh, St.~Andrews, KY16~9SS, Scotland, UK}

\author[0000-0001-9092-8619]{Jonathan Gelbord}
\affiliation{Spectral Sciences Inc., 4 Fourth Ave., Burlington, MA 01803, USA}

\author[0000-0003-0172-0854]{Erin Kara}
\affiliation{MIT Kavli Institute for Astrophysics and Space Research, Massachusetts Institute of Technology, Cambridge, MA 02139, USA}

\author[0000-0002-0380-0041]{Christian~Malacaria} 
\affiliation{International Space Science Institute, Hallerstrasse 6, 3012 Bern, Switzerland}

\author[0000-0001-8475-8027]{Jake A. Miller}
\affiliation{Department of Physics and Astronomy, Wayne State University, 666 W.\ Hancock St, Detroit, MI, 48201, USA}

\author[0000-0002-5872-6061]{James F. Steiner}
\affiliation{Center for Astrophysics, Harvard \& Smithsonian, 60 Garden St, Cambridge, MA 02138, USA}

\author[0000-0002-0118-2649]{Andrea Sanna}
\affiliation{Dipartimento di Fisica, Università degli Studi di Cagliari, SP Monserrato-Sestu km 0.7, 09042 Monserrato, Italy}

\begin{abstract}
The Seyfert 1 AGN Fairall~9 was targeted by NICER, Swift, and ground-based observatories for a $\sim1000$-day long reverberation mapping campaign. The following analysis of NICER spectra taken at a two-day cadence provides new insights into the structure and heating mechanisms of the central black hole environment. Observations of Fairall~9 with NICER and Swift revealed a strong relationship between the flux of the UV continuum and the X-ray soft excess, indicating the presence of a “warm” Comptonized corona which likely lies in the upper layers of the innermost accretion flow, serving as a second reprocessor between the “hot” X-ray corona and the accretion disk. The X-ray emission from the hot corona lacks sufficient energy and variability to power slow changes in the UV light curve on timescales of 30 days or longer, suggesting an intrinsic disk-driven variability process in the UV and soft X-rays. Fast variability in the UV on timescales shorter than 30 days can be explained through X-ray reprocessing, and the observed weak X-ray/UV correlation suggests that the corona changes dynamically throughout the campaign. 

\end{abstract}
\keywords{accretion, accretion disks --- 
black hole physics --- soft excess -- X-rays, UV, optical: individual (Fairall~9)}

\section{Introduction}
It is widely accepted that large galaxies contain a supermassive black hole (SMBH) near their center. Interaction with nearby material via accretion transforms the SMBH into an active galactic nucleus (AGN), characterized by extreme luminosity and variability in the X-ray through optical wavelengths. 
Quantifying the impact of AGNs on galaxy evolution requires an accurate description of the origin and energy budget of the X-ray source. Current telescopes cannot spatially resolve this region, so spectral and timing analysis techniques are used to infer its structure.

The Eddington ratio $\dot{m}_\mathrm{Edd}=L_\mathrm{Bol}/L_\mathrm{Edd}$ is a useful means to compare the accretion power of black holes across the mass scale. An AGN such as Fairall~9 ($M_\textit{BH}=2.6 \pm 0.6 \times 10^8 M_\odot$, \citealt{peterson04} and $z=0.047$, \citealt{Emmanoulopoulos11}), which accretes at a rate of $\dot{m}_\mathrm{Edd}= 0.06$ \citep{hagendone23}, is expected to form a geometrically thin, optically thick disk with a temperature profile ($T \propto R^{-3/4}$) which increases closer to the black hole \citep{Shakura1973}. The outer disk (optical/UV emission) spans $\sim$100--1000$R_\mathrm{G}$ ($R_\mathrm{G}=GM/c^2$) from the SMBH \citep{Frank02}. The inner disk may extend to the innermost circular stable orbit (ISCO) inward of $10 R_\mathrm{G}$ (\citealt{demarco13}; \citealt{kara16}), emitting in the extreme ultraviolet (EUV) and soft X-rays below 0.3 keV \citep{Laor_1997}, although intervening optically thick material prohibits direct observation. Hard X-rays (1--100 keV) are produced in a compact region within $\sim$10$R_\mathrm{G}$ (\citealt{Reis_2013}, \citealt{fabian14}, \citealt{ursini20}). The corresponding light crossing time between the hard X-ray and UV regions of Fairall~9 should be several hours, and a few days across the entire disk. 

In the reverberation picture, fluctuations in the UV and optical disk luminosity propagate outwards on the light crossing timescale and become delayed and smoothed, consistent with thermal reprocessing of emission from a central source (e.g., \citealt{Cackett2007}). Measuring the delay, or “lag,” in these fluctuations between different UV/optical bands provides a measurement of the light travel time between the disk radii (see \citealt{cackett21} for a recent review). This technique, known as continuum reverberation mapping, was recently used by \cite{Hernandez2020} to estimate the distance between the UV and optically emitting regions (1928--8700 \AA{}) in Fairall~9 of about 7 light days. On timescales of $\sim70$ days, slow variations in the optical are also shown to lead the UV, in the reverse direction of the reverberation lags \citep{Yao_2023}. A lag of $\sim10$ days in this direction was first reported in \cite{Hernandez2020}.

Reprocessing of X-ray emission is a possible driving mechanism for the short-term, outwardly propagating variability seen in the accretion disk. However, this has been challenged by significantly weaker correlation between variability in the X-rays/UV than in the UV/optical (e.g. \citealt{edelson19}, \citealt{Cackett_2023}). Part of the complication may lie in the fact that the soft X-ray emission originates from multiple components with ambiguous physical origins \citep{Lohfink_2016}, or a dynamic corona \citep{Panagiotou_2022_xraycorrel}. 

The hard X-ray spectrum follows a power-law shape, consistent with a ``hot corona" of electrons with $kT_e\sim$100 and a moderate optical depth of $\tau \sim 1-2$ (\citealt{fabian15}, \citealt{lubinski16}) which undergoes Comptonization with seed photons from the inner disk (\citealt{SUNYAEV79}; \citealt{Haardt_1991}). 
Although the hot corona is often assumed to be a spherical shell, reverberation measurements of multiple sources indicate a vertically extended component of variable height which may be the base of a failed jet (e.g. \citealt{wilkins13}, \citealt{wilkins16}, \citealt{Kara_2023}). 

Many AGN also exhibit a ``soft excess," X-ray emission below 2 keV in addition to the power law continuum from the hot corona. The soft excess can be described by a blackbody spectrum with a constant temperature of $kT=0.1-0.2$ keV, which is far too hot to be thermal emission from the disk (\citealt{gierlinski04}, \citealt{Crummy_2006}). Reverberation measurements on timescales of tens to hundreds of seconds show a contribution from hard X-rays which are reprocessed by the inner disk, becoming gravitationally redshifted and blurred by rotation (\citealt{zoghbi10}, \citealt{wilkins13}). This has been confirmed for numerous AGN, and can be used to constrain the radius of the inner disk and hot corona (e.g. \citealt{Emmanoulopoulos14}, \citealt{cackett14}, \citealt{Hancock22}).

However, numerous studies demonstrate that variability in the soft excess on longer timescales cannot be explained entirely by reflection, either through spectral modeling (e.g. \citealt{boissay16}, \citealt{porquet18}) or time variability (e.g. \citealt{Tortosa_2023}, \citealt{Zoghbi_2023}). Often, the soft excess is more closely linked to variability in the luminosity of the UV/optical disk, likely due to changes in the accretion rate (e.g. \citealt{matt14}, \citealt{Mehdipour2015}, \citealt{mahmoud20}, \citealt{mahmoud22}, \citealt{middei23}, \citealt{Mehdipour_2023}).

These results suggest that the dominant contribution to the soft excess may be the X-ray tail of a larger “warm” corona, an optically thick ($\tau \sim10-20$) Comptonization region with electron temperatures of $kT\approx1$ keV, which would emit mostly in the EUV (\citealt{mehdipour11}, \citealt{Done12}). If the blackbody-emitting disk is truncated before reaching the ISCO, the warm corona may be the innermost part of the accretion flow. Alternatively, the warm corona may be a vertically extended layer above and below the inner-disk plane, which acts as the source of seed photons for Comptonization (\citealt{Petrucci_2018}, \citealt{kubota18}, \citealt{Ballantyne_2020}). In a survey of AGN spectra using eROSITA, 23 out of 29 AGN with a strong soft excess preferred a warm corona model, while the remaining six were better characterized by a blurred relativistic reflection model \citep{waddell_2023}. This suggests that while both of these features may exist in a given AGN, there are internal conditions present which can cause either feature to dominate the spectrum.

The relationship between long-term variability in the disk and the warm corona, in conjunction with disk reverberation results, suggests that the soft excess may be the elusive source which drives disk reprocessing. Confirming this requires luminosity measurements of the soft excess on timescales of a few days, the same rate at which short term variability in the inner accretion disk occurs. The contribution from the hot corona must also be measured, separately from the soft excess and over the same energy range. This requires robust X-ray spectral modeling of the target with a cadence of at least once every two days for a black hole with $M_\textit{BH} \approx 10^8 M_\odot$.

\begin{figure*}[tp]
    \centering
    \includegraphics[width=0.9\textwidth]{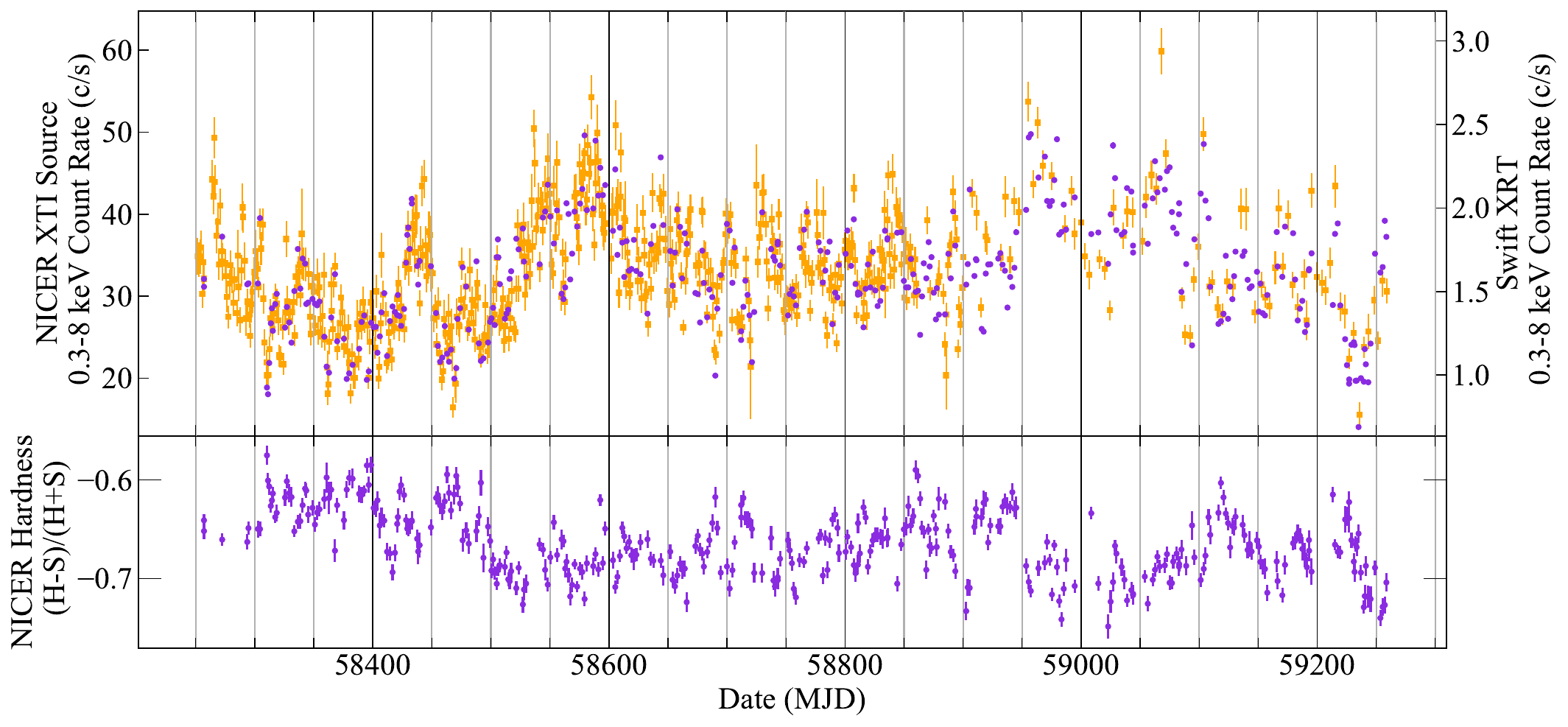} 
\caption{\textbf{\textit{Top}}: Light curve of Fairall~9 using NICER XTI (purple) and Swift XRT (orange) from 0.3--8 keV. The axes are rescaled to account for NICER's $20\times$ greater sensitivity than Swift. NICER source count rates are measured by subtracting the contribution of best-fitting SCORPEON background model. \textbf{\textit{Bottom}}: Hardness ratio of NICER observations, which demonstrates variability in the observed spectral shape. The H band covers 1.5--8 keV, while the S band covers 0.3--1.5 keV.}  
\label{fig:lcCompSwiftNICERXrays}
\end{figure*}

Swift can measure the broadband flux variability of Fairall~9 in the optical, UV, and X-rays at the necessary cadence, but cannot produce X-ray spectra sufficient for this experiment due to its low count rate sensitivity ($\sim$2 c/s, see Figure \ref{fig:lcCompSwiftNICERXrays}). Only NICER is capable of doing this. Individual 1 ks observations of Fairall~9 ($\sim$30 c/s) are sufficient to develop detailed spectral models of AGN, which was also demonstrated in a similar experiment on Mrk 817 ($\sim 1-10$ c/s) by \cite{partington23}. 

To understand the nature of the soft excess in Fairall~9 and its role in X-ray reprocessing, we will compare variability in the fluxes of the X-ray power law continuum, the X-ray soft excess, and the UV continuum. Our cross-correlation analysis will determine if trends in the soft excess and the UV are closely related to the X-ray emission from the hot corona, consistent with X-ray reprocessing scenario. To test if the X-rays can power the observed UV variability on timescales of days to months, we will estimate the energy released by the hot X-ray corona and the accretion disk. We will also search for slow variability intrinsic to the disk in both the soft excess and the UV light curves. If the soft excess originates from a warm corona above the accretion disk, we would expect to see common trends in both the soft excess and the UV light curves which are not observed in the X-ray power law continuum. This approach will yield new information about the location and geometry of the soft excess source.

\section{Observations and Data Reduction}
\label{sec:Data}
Observations of Fairall~9 were taken with the NICER XTI from 18 May 2018 to 13 Feb 2021 (MJD 58256--59258) at a cadence of approximately two days, as part of the observations dedicated to the Observatory Science Working Group (Target IDs: 110002, 210002, 310002). The data were processed using the \textsc{HEASOFT} version 6.32.1 (\citealt{Blackburn_1995}) and \textsc{CALDB} version \textsc{xti20221001}. The event files were screened using \textsc{NICERL2} with standard settings, including the exclusion of anomalous FPMs and MPUs with \textsc{niautoscreen}. Spectra and response files were produced for each observation using \textsc{NICERL3-spect}. We excluded observations with a total filtered exposure $<$ 300 s to ensure sufficient quality for spectral modeling, resulting in 404 epochs. 

The total count-rate spectra are background-dominated above 8 keV, with a median 8--10 keV count rate of 0.13 c s$^{-1}$ across all epochs. Rare epochs with extreme background interference from high energy particles reduce the efficacy of the NICER background estimators (see \citealt{partington23}) and the \textsc{SCORPEON} background model\footnote{\url{https://heasarc.gsfc.nasa.gov/docs/nicer/analysis_threads/scorpeon-overview}} (see Appendix~\ref{sec:bgappendix}). These background flares can dominate the spectrum, preventing accurate measurement of the power law emission from Fairall~9. We exclude 21 epochs which exhibit these flares, characterized by a total source plus background count rate exceeding 3.15 c s$^{-1}$ in the 8--10 keV band (the top 5\% of all epochs). This leaves 383 epochs in our final sample. 

Source (background-subtracted) count rates are calculated using the \textsc{SCORPEON} background model and the source model in Section~\ref{sec:tiedsoftexcess}. Light curves are presented in Figure~\ref{fig:lcCompSwiftNICERXrays}, with contemporaneous \textit{Swift} XRT monitoring for comparison. NICER source count rates range from 14.0--49.8 counts s$^{-1}$ with a median of 32 counts s$^{-1}$.

We also generate spectra and count rate light curves of contemporaneous Swift XRT observations using the XRT Product Builder (\citealt{evans07}, \citealt{evans09}) for comparison with NICER. The shapes of the 0.3--8 keV light curves shown in Figure~\ref{fig:lcCompSwiftNICERXrays} are consistent between the two instruments. We detect a difference in the spectral shape measured by Swift XRT and NICER XTI, indicating an offset in calibration discussed in Appendix~\ref{sec:swiftappendix}. Swift UVOT light curves in the UVW2 band are produced following the methodology of \cite{Hernandez2020}.

\section{Analysis}
\label{sec:Analysis}
\subsection{Source and Background Spectral Model}
\label{sec:spectralmodel}
Spectra from each NICER epoch are fit in XSPEC \textsc{v12.13.1} \citep{arnaud96}. The total count rate spectrum, including both the source and background, are binned using the \textsc{FTOOLS} module \textsc{FTGROUPPHA} according to the optimal binning scheme developed by \cite{kaastra16}, which is based on the instrument's resolution. Adjacent bins are also combined to ensure a minimum of 25 counts per bin. The spectral shape of the source and background are fit simultaneously from 0.22--15 keV in XSPEC using $\chi^2$ minimization. 

\begin{figure*}[tp]
    \centering
    \includegraphics[width=0.9\textwidth]{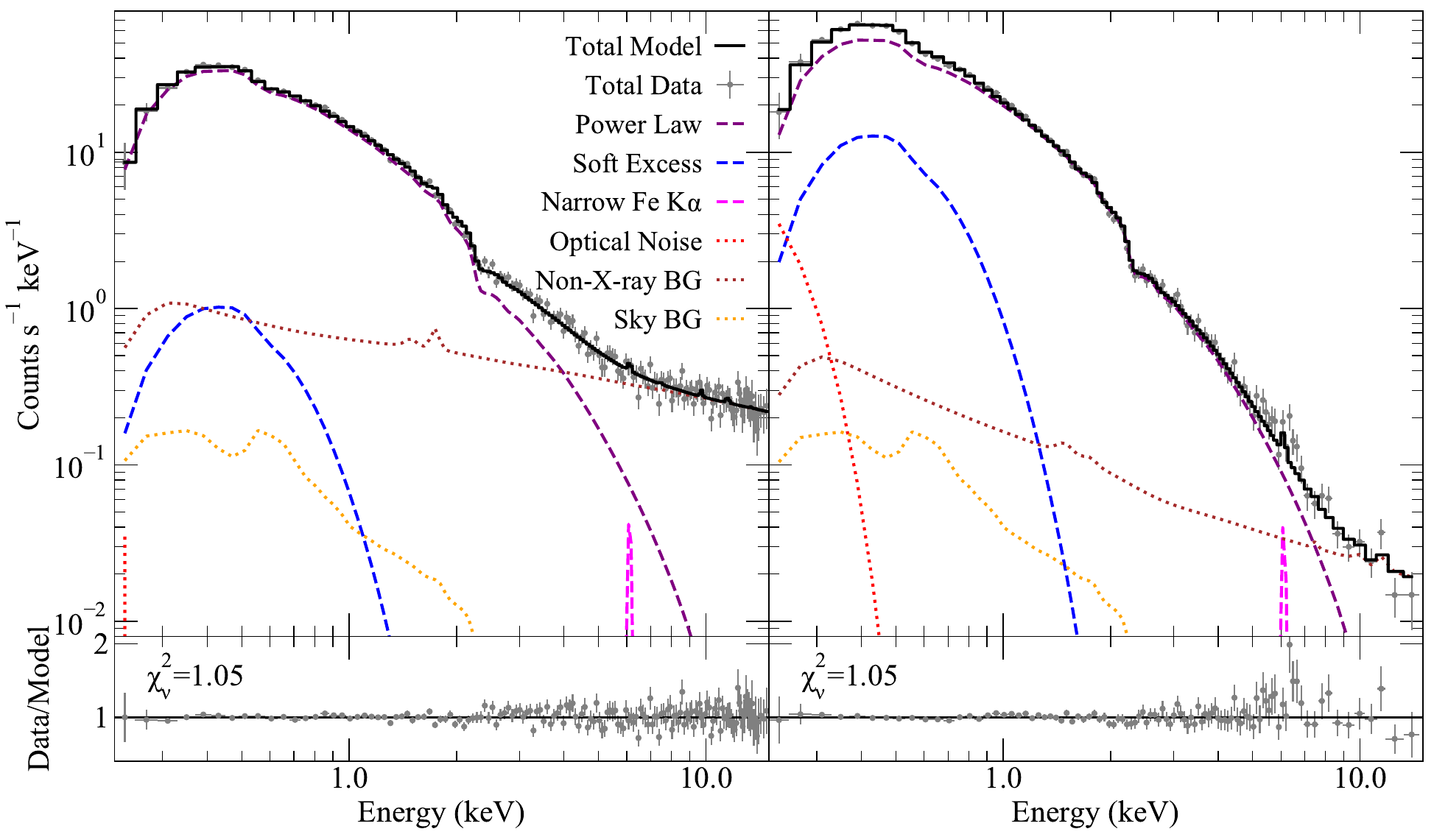} 
\caption{Photon count spectra from two NICER observations which demonstrate a change in the X-ray soft excess, occurring independently from the hard X-ray power law state. The total data (source and background) are shown as grey points, with the best-fit model from Section \ref{sec:tiedsoftexcess} in black. The data-to-model ratios for each bin are shown as grey points below. The source model (dashed lines) contains a power law continuum (purple), a blackbody soft excess (blue), and a narrow Gaussian representing the Fe K$\alpha$ line (pink). The background model (dotted lines) contains a Gaussian optical noise peak (red), the non-X-ray background (brown) and the sky X-ray background (yellow). \textbf{\textit{Left}}: This observation was taken on MJD 58400 when the Swift UVW2 flux was near minimum (ObsID 1100020169). The X-ray power law has flux $\log(\Phi_{\mathrm{PL}})=-10.347\pm0.004$ (all fluxes are reported in units of erg~cm$^{-2}$~s$^{-1}$) and $\Gamma=2.11\pm0.01$. The soft excess flux is $\log(\Phi_{\mathrm{SE}})=-12.364\pm0.008$. \textbf{\textit{Right}}: An observation from MJD 59060 taken near Swift UVW2 maximum (ObsID 3100020364), with $\log(\Phi_{\mathrm{PL}})=-10.200\pm0.003$ and $\Gamma=2.23\pm0.01$. The soft excess flux is $\log(\Phi_{\mathrm{SE}})=-11.260\pm0.008$.}
\label{fig:spec_comp}
\end{figure*}

The \textsc{SCORPEON} background model is calibrated using NICER observations of the sky in regions with no X-ray sources. It contains variable components which characterize the spectral shape of the cosmic, Galactic, and solar X-ray background and the high energy particle background. Optical noise is characterized with a Gaussian curve centered at 110 eV with variable height and width. This accounts for ``optical loading," in which optical light produces a current in the Silicon Drift Detectors, creating a false soft X-ray signal. Optical noise is typically present in the spectrum below 0.2 keV, outside of the 0.3--10 keV source sensitivity band for NICER. Extreme optical noise present at up to 0.5 keV can occur when the target is observed at a low sun angle, accounted for in the model by the variable width of the Gaussian (see Fig.~\ref{fig:spec_comp}). All parameters for the sky background are fixed at their respective values calculated by the SCORPEON estimator, and are constant throughout the campaign.

The source model includes Galactic absorption using \textsc{tbabs} \citep{Wilms_2000} with a fixed value of $N_H=2.85 \times 10^{20}$ $\text{cm}^{-2}$, calculated using the \textsc{FTOOL} \textsc{nh}. The X-ray continuum is modeled using a power law, and the unabsorbed fluxes of each emission component are calculated from 0.3--8 keV using \textsc{cflux}. We test the relativistic reflection model \textsc{relxill} to describe the soft excess and reflection spectrum. However, in many epochs the background and source fluxes are comparable in the energy range of the broad Fe K$\alpha$ line (see Fig.~\ref{fig:spec_comp}), making it difficult to constrain the shape of this feature. Thus, any fit with \textsc{relxill} is largely based on the shape of soft excess. The spectra are insufficient to constrain model parameters such as spin, inclination, and disk ionization. This motivates our decision to use the phenomenological \textsc{blackbody} model, which has a comparable spectral shape below 2 keV. 

During some observations with a bright optical background from the Sun or strong high energy particle interference, the blackbody temperature $kT$ shifts to the boundaries of the allowed range ($kT=0.05$--$0.2$ keV) in order to fit the background. However, during all observations with a low background state, $kT$ remains between 0.10--0.12 keV. This small variation is consistent with observations of other AGN (e.g. \citealt{Crummy_2006}). This motivates our decision to first fit each spectrum with $kT$ as a free parameter, and then fix the parameter to the median value across all epochs of $kT=0.106$ keV. 

We model the narrow Fe K$\alpha$ line caused by distant reflection using a Gaussian with zero width, fixed at a line energy of 6.1 keV given a rest energy of 6.4 keV and $z=0.047$ \citep{Emmanoulopoulos11}. Observations with short exposures or bright power law flux are insensitive to the flux of the narrow Fe K$\alpha$ line, so we first allow this parameter to be free, and then fix the flux to the median ($\Phi_{\mathrm{K}\alpha}=10^{-12.67}$ erg~cm$^{-2}$~s$^{-1}$) for the final model. Since this emission is thought to originate from distant gas in the torus, which is not expected to vary significantly on the timescale of our campaign, we use the same value for each epoch. 

The final model in XSPEC reads: \textsc{tbabs*(cflux*
powerlaw+cflux*blackbody+cflux*gaussian)}. The flux of the power law is always free, and allowed parameter ranges are shown in Table~\ref{table:modelparams}. NICER spectra are not sensitive to the cutoff energy of the power law which exceeds $500$ keV \citep{Lohfink_2016}, and it is not needed to fit our data, so it is not included in our model. The blackbody flux and the power law index $\Gamma$ are strongly degenerate (see Fig.~\ref{fig:cornerplot}), so we test alternative models with each of these parameters held fixed as described in Sections~\ref{sec:fixedgamma} and~\ref{sec:tiedsoftexcess}.

\begin{deluxetable}{ccc}
\tablewidth{0pt}
\tablecaption{\label{table:modelparams}Source Model Parameters}
\tablehead{
\colhead{Component} & \colhead{Parameter} & \colhead{Allowed Range}}
\startdata
\textsc{tbabs} & $N_\mathrm{H}$ ($10^{22}$cm$^{-2}$) & 0.0285 \tablenotemark{a} \\
\hline
\textsc{cflux} & $\log(\Phi_{\mathrm{PL}})$ (erg$\cdot$cm$^2\cdot$s$^{-1}$) & $[-16,-9]$\\
\textsc{powerlaw} & $\Gamma$ & $[1.6,3]$ \tablenotemark{b}\\
\hline
\textsc{cflux} & $\log(\Phi_{\mathrm{SE}})$ (erg$\cdot$cm$^2\cdot$s$^{-1}$) & $[-16,-9]$ \tablenotemark{c}\\
\textsc{blackbody} &$kT$ (keV) &  0.106 \tablenotemark{a} \\
\hline
\textsc{cflux} & $\log(\Phi_{\mathrm{K}\alpha})$ (erg$\cdot$cm$^2\cdot$s$^{-1}$) & $-12.67$ \tablenotemark{a}\\
\textsc{gaussian} & line energy (keV) &  6.1 \tablenotemark{a} \\
 & line width (keV) &  0 \tablenotemark{a} \\
\hline
\enddata
\tablenotetext{a}{Always fixed (see Section~\ref{sec:spectralmodel}).}
\tablenotetext{b}{Fixed to $\Gamma=2.15$ in Section~\ref{sec:fixedgamma}.}
\tablenotetext{c}{Tied to UVW2 flux density in Section~\ref{sec:tiedsoftexcess}.}
\end{deluxetable}

\subsection{Fixed Power Law Index Model}
\label{sec:fixedgamma}
\begin{figure}[t]
  \centering
  \includegraphics[trim={.9cm 0 0 0},clip,width=\columnwidth]{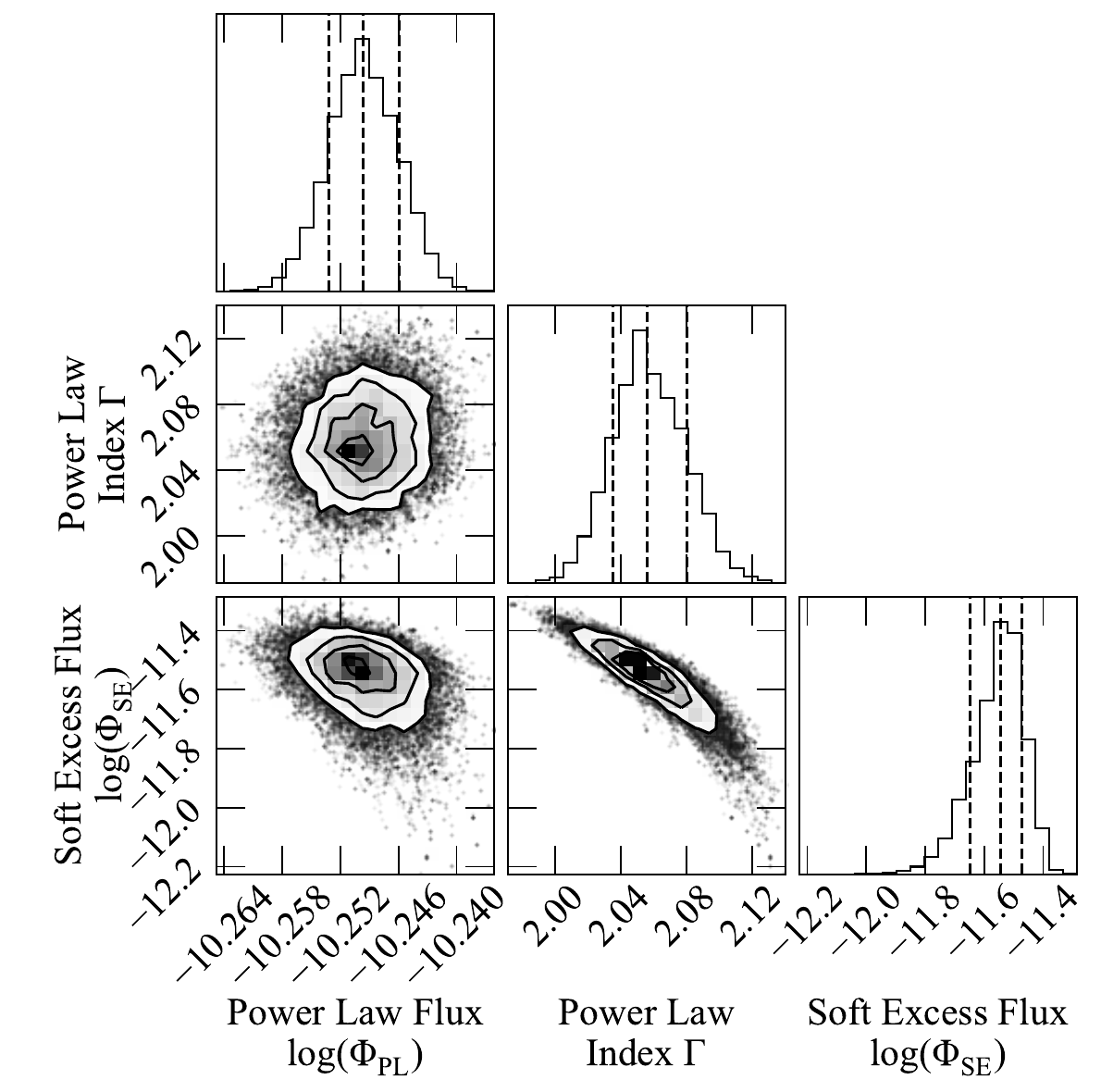}
  \caption{Corner plot showing the parameter distributions for the model in Table~\ref{table:modelparams} applied to the observation on MJD 58256. Both the power law index $\Gamma$ and the blackbody soft excess flux $\log(\Phi_{\mathrm{SE}})$ are free parameters, and are fit using the MCMC method with 100,000 steps and an initial burn-in of 20,000 steps. Dashed lines on the histograms represent the lower $1\sigma$ confidence interval, the median, and the upper $1\sigma$ interval. Contours represent $0.5\sigma$, $1\sigma$, $1.5\sigma$, and $2\sigma$. A clear degeneracy between $\Gamma$ and $\log(\Phi_{\mathrm{SE}})$ is evident in the central bottom panel, given the shape of the distribution.}
  \label{fig:cornerplot}
\end{figure}

\begin{figure*}[tp]
    \centering
    \includegraphics[width=0.9\textwidth]{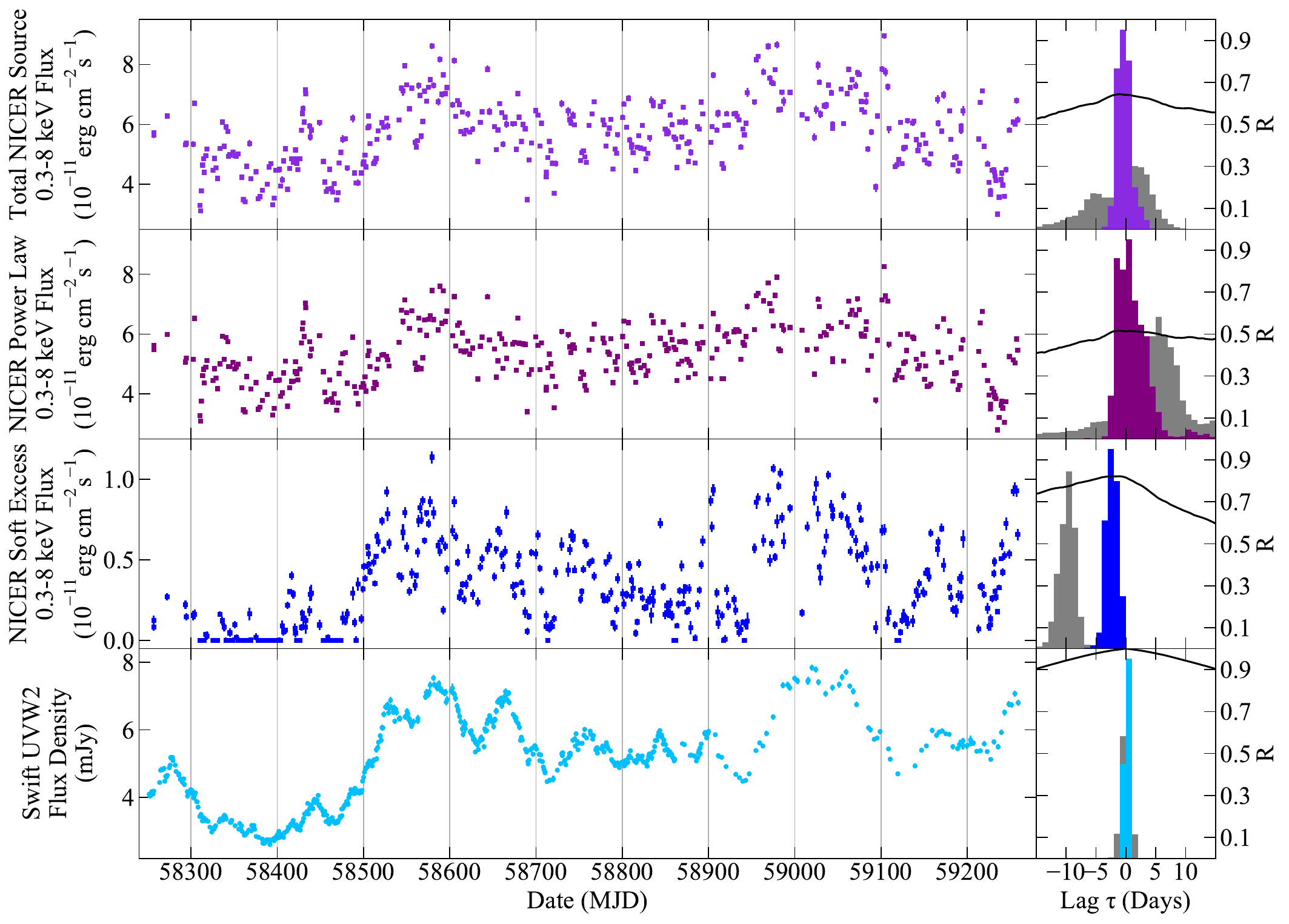}
\caption{\textbf{\textit{Left}}: Light curves of the best-fit flux values of the fixed $\Gamma=2.15$ model in Section~\ref{sec:fixedgamma} for each NICER spectrum, in descending order: total source flux (light purple), power law continuum flux ($\log(\Phi_{\mathrm{PL}})$, dark purple), blackbody soft excess flux ($\log(\Phi_{\mathrm{SE}})$, dark blue), plus the Swift UVW2 flux density ($\phi_{\nu,\mathrm{UVW2}}$, light blue). Near UVW2 minimum (MJD 58300--58500), fixing $\Gamma$ produces model fluxes higher than those observed in the soft X-rays, causing the soft excess flux to be pegged at the minimum value of $10^{-16}$ erg~cm$^{-2}$~s$^{-1}$ (see Table~\ref{table:modelparams}). \textbf{\textit{Right}}: Cross-correlation functions relative to the UVW2 reference band (solid line), with histograms representing the probability distributions of the peak (purple) and centroid (grey) lag values.}  
\label{fig:fixedgammalightcurves}
\end{figure*}

To address the degeneracy between the power law index, which describes the slope of the emission from the hot X-ray corona, and the flux of the soft excess (see Figure~\ref{fig:cornerplot}), we fix the power law index to $\Gamma=2.15$. This is the median value from the analysis in Section~\ref{sec:spectralmodel} over the 383 epochs. Light curves with modeled fluxes for each epoch are shown in Figure~\ref{fig:fixedgammalightcurves} and Table~\ref{table:fixedgammaparam}. The total $\Sigma(\chi^2)$, summed across spectra from all 383 epochs, is 58,321 for 40177 degrees of freedom ($\chi^2_\nu=1.45$).

We note that this model is not a satisfactory physical description of the X-ray power law variability. For instance, when Fairall~9 is faintest in X-rays and UVW2 from MJD 58300--58500, the fit at soft energies is often unsuitable for $\Gamma=2.15$ since the soft excess fluxes are unconstrained in many epochs (see Fig.~\ref{fig:fixedgammalightcurves}). If $\Gamma$ is left free, the residuals diminish and the best-fit value approaches $\Gamma=1.9$. This is characteristic of the softer-when-brighter behavior of AGN \citep{Magdziarz_1998}. These issues motivate our alternative test with a soft excess flux set as a function of the UVW2 flux in Section \ref{sec:tiedsoftexcess}.  However, the fixed $\Gamma=2.15$ test does provide insight into the separate behavior of the soft and hard X-ray spectral components.

\begin{deluxetable*}{cccccccc}
\tablewidth{0pt}
\tablecaption{\label{table:fixedgammaparam}Spectral Analysis Results: NICER background-subtracted count rate, hardness ratio (H-S)/(H+S), and model component fluxes for the fixed $\Gamma=2.15$ model in Section~\ref{sec:fixedgamma}. Fluxes are calculated from 0.3--8 keV. Uncertainties corresponding to the 68\% confidence interval ($1\sigma$) are reported for each parameter. The full table is published in machine-readable format.}
\tablehead{
\colhead{Obs. Date} &
\colhead{Count Rate} &
\colhead{Hardness Ratio} &
\colhead{Total Source}&
\colhead{Power Law} &
\colhead{Soft Excess} &
\colhead{Best fit} & 
\colhead{Degrees}\\ 
\colhead{(MJD)} &
\colhead{(counts~s$^{-1}$)} &
\colhead{H=1.5--8 keV}&
\colhead{$\log(\Phi_{\mathrm{PL}}+\Phi_{\mathrm{SE}}+\Phi_{\mathrm{K}\alpha})$} &
\colhead{$\log(\Phi_{\mathrm{PL}})$} &
\colhead{$\log(\Phi_{\mathrm{SE}})$} &
\colhead{$\chi^2_{\nu}$} &
\colhead{of}\\ 
\colhead{} &
\colhead{0.3-8 keV} &
\colhead{S=0.3--1.5 keV} &
\colhead{(erg~cm$^{-2}$~s$^{-1}$)} &
\colhead{(erg~cm$^{-2}$~s$^{-1}$)} &
\colhead{(erg~cm$^{-2}$~s$^{-1}$)} &
\colhead{} &
\colhead{Freedom} 
}
\startdata
58256.95 & $31.91\pm0.17$ & $-0.650\pm0.006$ & $-10.242\pm0.005$ & $-10.250\pm0.003$ & $-12.07\substack{+0.09\\ -0.14}$ & $131.87$ & $126$ \\ 
58257.00 & $31.19\pm0.22$ & $-0.654\pm0.009$ & $-10.249\substack{+0.006\\ -0.008}$ & $-10.260\substack{+0.004\\ -0.006}$ & $-11.91\substack{+0.10\\ -0.12}$ & $110.87$ & $116$ \\ 
58272.38 & $37.26\pm0.20$ & $-0.660\pm0.006$ & $-10.202\pm0.004$ & $-10.223\pm0.003$ & $-11.57\substack{+0.03\\ -0.04}$ & $392.05$ & $158$ \\ 
58293.62 & $31.51\pm0.20$ & $-0.659\pm0.008$ & $-10.275\substack{+0.006\\ -0.007}$ & $-10.295\pm0.005$ & $-11.65\pm0.05$ & $167.40$ & $157$ \\ 
58294.65 & $31.53\pm0.18$ & $-0.652\pm0.007$ & $-10.270\pm0.005$ & $-10.284\substack{+0.004\\ -0.003}$ & $-11.83\substack{+0.06\\ -0.07}$ & $134.53$ & $98$ \\ 
... & ... & ... & ... & ... & ... & ... & ...
\enddata
\end{deluxetable*}

\subsection{Cross-Correlation Analysis}
We test the correlation between variability in the disk and X-ray emission components using the Swift UVW2 band as a reference. A detailed description of the Swift data reduction and analysis with other UVOT bands will be presented in Edelson et al., submitted, and an analysis of the first year of Swift observations is presented in \cite{Hernandez2020}. We use the Interpolated Cross Correlation Function (ICCF), which shifts each X-ray light curve relative to the UVW2 curve by a lag $\tau$ and measures the Pearson correlation coefficient $R(\tau)$ between the interpolated light curves for each $\tau$ \citep{Peterson_1998}. Lags within a range of $\tau=[-100,100]$ days are tested using a timestep of 0.2 days. Uncertainties are calculated using the Random Subset Selection (RSS) method and the Flux Randomization (FR) Markov Chain Monte Carlo method with 25,000 realizations \citep{Sun_2018}. The reported ``centroid lag" is defined to be the mean of the centroid values calculated by the FR/RSS methods for $R(\tau) > 0.8R(\tau)_\mathrm{Peak}$. 

Figure~\ref{fig:fixedgammalightcurves} shows that the peak and centroid lag values sometimes differ significantly, associated with a relatively flat ICCF. This is likely caused by the overall slow variability present in the light curves, such as from MJD 58475--58525, which tilts the  ICCF enough to affect its centroid more than its peak. Given the consistent sampling by both Swift and NICER and our goal of measuring time lags between fast variability, we proceed using the peak  ICCF values for our analysis. We report $R_\mathrm{peak}$ and $\tau_\mathrm{peak}$ as the median values of the respective probability distribution functions for each  ICCF, with uncertainty ranges representing the central 68\% of values ($1\sigma$). These are presented in Table~\ref{table:peaklags}.

\begin{deluxetable*}{ccccc}
\tablewidth{0pt}
\tablecaption{\label{table:peaklags}X-ray/UV Correlation and Lags: Peak correlation ($R$) and lag ($\tau$) values for NICER spectral component light curves and the power law index ($\Gamma$). We use the Swift UVW2 flux density as a reference band, and negative lags indicate that the X-ray variability leads the UV. Uncertainty ranges corresponding to the central 68\% of values ($1\sigma$) are reported for each lag measurement.}
\tablehead{
\colhead{} &
\multicolumn{2}{c}{Fixed $\Gamma=2.15$}  &  \multicolumn{2}{c}{UVW2-linked $\log(\Phi_{\mathrm{SE}})$} \\
\colhead{Parameter} & 
\colhead{$R_\mathrm{peak}$} & 
\colhead{$\tau_\mathrm{peak}$ (days)} &
\colhead{$R_\mathrm{peak}$} & 
\colhead{$\tau_\mathrm{peaks}$ (days)}
}
\startdata
Total Flux& $0.64$&$-0.6\substack{+1.4\\ -0.6}$& $0.60$&$0.4\pm1.6$ \\ 
Power Law Flux $\log(\Phi_{\mathrm{PL}})$& $0.51$&$0.8\substack{+2.8\\ -2.0}$& $0.50$&$0.6\substack{+2.6\\ -2.0}$ \\ 
Power Law Index $\Gamma$& -- \tablenotemark{a}& -- \tablenotemark{a}& $0.54$&$-3.2\substack{+1.0\\ -2.2}$ \\ 
Soft Excess Flux $\log(\Phi_{\mathrm{SE}})$& $0.82$&$-2.4\substack{+1.4\\ -1.0}$& $1.00$ \tablenotemark{b}&$-1.2\substack{+0.3\\ -0.1}$ \tablenotemark{c} \\ 
\hline
\enddata
\tablenotetext{a}{Fixed to $\Gamma=2.15$ in Section~\ref{sec:fixedgamma}.}
\tablenotetext{b}{Tied to UVW2 flux density in Section~\ref{sec:tiedsoftexcess}.}
\tablenotetext{c}{Calculated from the grid search in Section~\ref{sec:tiedsoftexcess}.}
\end{deluxetable*}

For the light curve of the total modeled 0.3--8 keV flux, $R_{\mathrm{Peak}}=0.64$. $R_{\mathrm{Peak}}=0.51$ for the isolated power law component. Both lags are consistent with zero to $1\sigma$. The relatively low correlation coefficients arise mainly from fast X-ray variations that are either absent or washed out by time-smearing in the UVW2 light curve. The strongest X-ray/UV correlation observed in this source to date comes from the soft excess component, with $R_{\mathrm{Peak}}=0.82$ and $\tau=-2.4\substack{+1.4\\ -1.0}$ days, indicating that the soft excess variability leads the UVW2.

\subsection{UVW2-Linked Soft Excess Model}
\label{sec:tiedsoftexcess}
The strong correlation between the light curves of the UVW2 flux density and X-ray soft excess flux in the fixed $\Gamma=2.15$ model motivates an alternative test allowing for natural variability in the power law index. We assume that the shape of the light curve in the soft excess directly matches the UVW2, with fluxes offset by a fixed amount, scaled by a factor $A$, and shifted in time by a lag $\tau$ using Equation~(\ref{eq:SEUVfluxscal}). Here, a negative lag indicates that the X-ray soft excess leads the UVW2, per the convention in \cite{Hernandez2020}. This assumption is similar to the one made in PyROA \citep{Donnan_2021}, which is commonly used for light curve modeling in reverberation mapping experiments.

While this method may introduce a bias in the shape of the soft excess light curve, it is a necessary step towards obtaining a physically meaningful measurement of the X-ray spectral variability. As discussed in Section~\ref{sec:spectralmodel}, the degeneracy between $\Gamma$ and the soft excess flux ($\Phi_{\mathrm{SE}}$) results in poorly-constrained values for each parameter if both are left free during fitting. Additionally, the ICCF between $\Phi_{\mathrm{SE}}$ and the UVW2 flux density  is flat if both $\Gamma$ and $\Phi_{\mathrm{SE}}$ are free, resulting in X-ray/UV lags in all bands that are consistent with zero with uncertainties of $\sim2$ days. This motivates our decision to pursue a function that describes $\Phi_{\mathrm{SE}}$.

We first linearly interpolate the UVW2 flux density values ($\phi_{\nu,\mathrm{UVW2}}$, note that we use different symbols for flux and flux density) to produce a light curve for any given time $t$ of $\phi_{\nu,\mathrm{UVW2}}(t)$. We then calculate the corresponding flux value of the X-ray soft excess $\Phi_{\mathrm{SE}}(t)$ at that time, introducing a common time-shift of $\tau$ and a common multiplicative factor of $A$. We use a fixed offset ($\phi_{\nu,\mathrm{min,UVW2}}$) to equate the minimum UVW2 flux density on MJD 58391 ($2.61 \pm 0.05$ mJy) to a soft excess flux of zero. This is based on the non-significant detection of the soft excess in the NICER spectra during UVW2 minimum (see Fig.~\ref{fig:spec_comp}). The calculation of $\Phi_{\mathrm{SE}}(t)$ reads 

\begin{equation}
\label{eq:SEUVfluxscal}
    \Phi_{\mathrm{SE}}(t) = A \left(\phi_{\nu,\mathrm{UVW2}}(t-\tau) - \phi_{\nu,\mathrm{min,UVW2}} \right).
\end{equation}

We search for best-fitting parameters $\tau$ and $A$ to describe the relationship between the UVW2 and soft excess light curves by generating a grid of soft excess fluxes for each observation, using pairs of $\tau$ and $A$. The grid search is done in lieu of fitting all 383 epochs simultaneously, which cannot manageably be done in XSPEC. We instead calculate $\Phi_{\mathrm{SE}}$ for each NICER epoch for a given $\tau$ and $A$, which are set as fixed parameters in XSPEC. The flux $\Phi_{\mathrm{PL}}$ and $\Gamma$ of the power law component are allowed to vary in the model. We test a range of $A$ from $0.46$ to $4.1\times 10^{-12}$ erg~cm$^{-2}$~s$^{-1}$~mJy$^{-1}$ and $\tau$ from $-$15 to 15 days. 

For computational ease, we initially fix the SCORPEON background model parameters to their respective best-fit values from the fixed $\Gamma=2.15$ analysis in Section~\ref{sec:fixedgamma}. We thus leave only the power law flux $\Phi_{\mathrm{PL}}$ and spectral index $\Gamma$ free in our $\Sigma(\chi^2)$ comparison when calculating $\tau$ and $A$. We then release the fixed background parameters in a final fit to estimate parameter uncertainties for our final light curves. Note that the mean difference in the 0.3--8 keV background count rate between the results here and in Section~\ref{sec:fixedgamma} is 0.002 c/s, compared to a mean source count rate of 32.6 c/s, so this choice should not affect our results. 

\begin{figure}[t]
  \centering
  \includegraphics[width=\columnwidth]{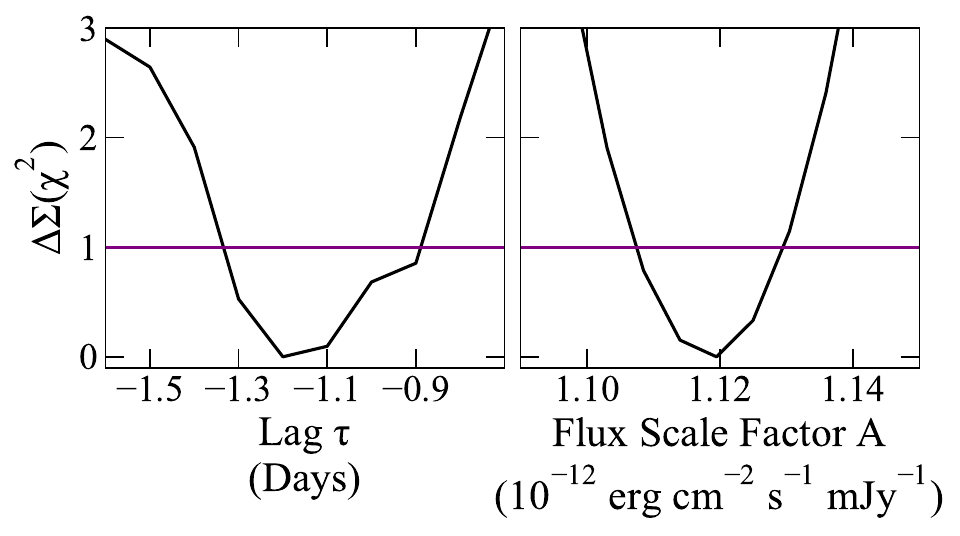}
  \caption{Changes in the best-fit $\chi^2$ value of the ``UVW2-linked soft excess" model, created by shifting the soft excess light curve by lag ($\tau$, left) and UVW2/soft excess flux scaling factor ($A$, right). The $\chi^2$ statistic is summed across all 383 epochs for each value of $\tau$ and $A$. Purple lines highlight the $\Delta\Sigma(\chi^2)$=1, and intersections with the black line represent the limits of the 68\% confidence interval ($1\sigma$) for $\tau$ and $A$.}

  \label{fig:1dgridsearch}
\end{figure}

\begin{figure*}[t]
  \centering
  \includegraphics[width=\textwidth]{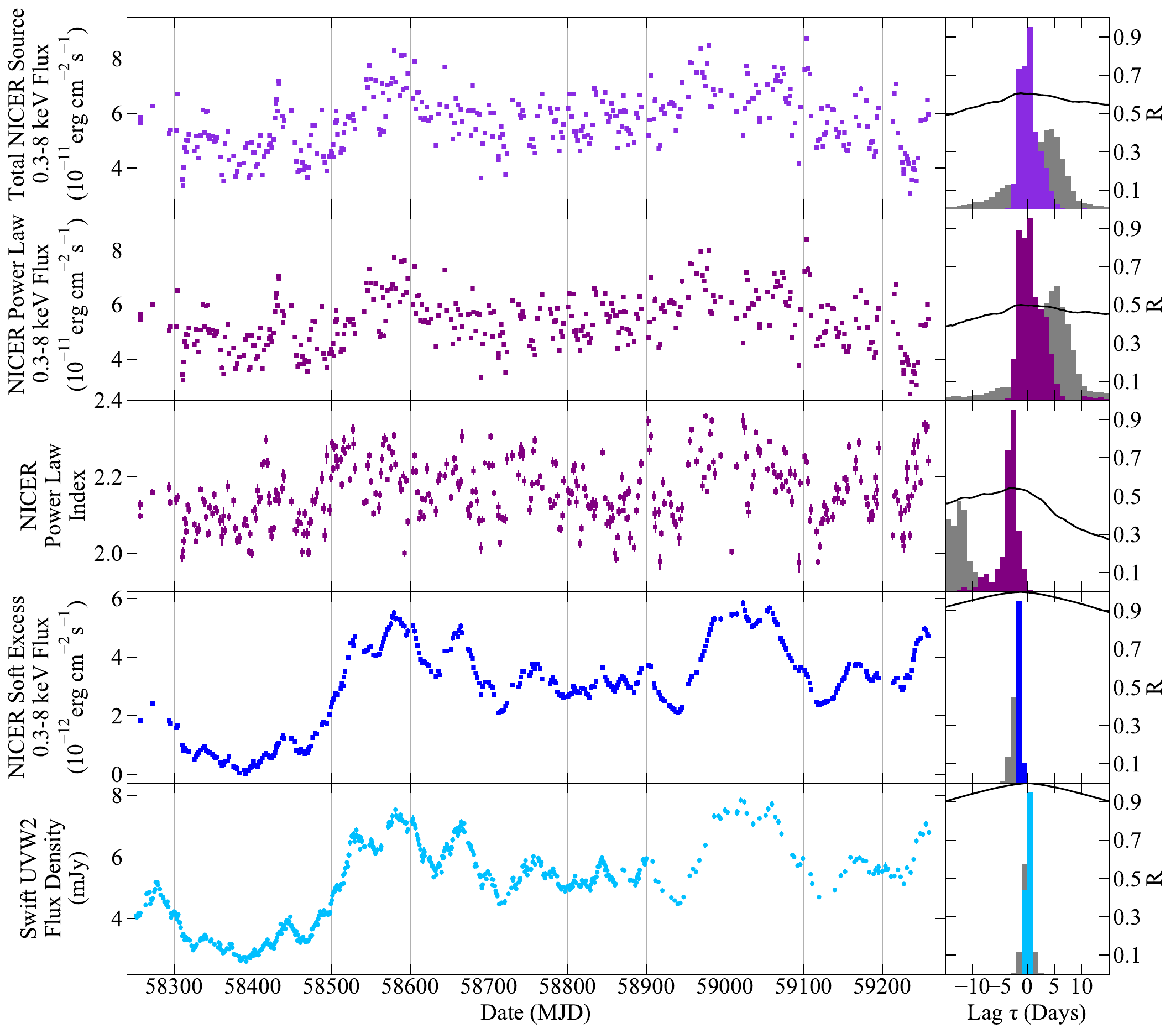}
  \caption{Same as Fig.~\ref{fig:fixedgammalightcurves}, but for the best-fitting UVW2-linked soft excess model in Section~\ref{sec:tiedsoftexcess}. The central panel shows the power law index $\Gamma$ which represents the spectral shape of the hot X-ray corona, and was a fixed parameter in the light curves of Fig.~\ref{fig:fixedgammalightcurves}. Here, the soft excess flux curve (Panel 4) is determined by Equation~\ref{eq:SEUVfluxscal}, with best-fitting values for $A$ and $\tau$ shown in Fig.~\ref{fig:1dgridsearch}, and is held fixed during spectral fitting.}
  \label{fig:UVW2linklightcurves}
\end{figure*}

The best-fit $\chi^2$ statistic summed across all 383 epochs is $\Sigma(\chi^2)=53,614$ for 40,177 degrees of freedom ($\chi^2_\nu=1.33$), using $\Phi_{\mathrm{SE}}(t)$ generated from $\tau=-1.2 \substack{+0.3\\ -0.1}$ days and $A=1.12 \pm0.02 \times 10^{-12}$ erg~cm$^{-2}$~s$^{-1}$~mJy$^{-1}$. Figure~\ref{fig:1dgridsearch} shows $\chi^2$ minimised along the nuisance parameters of the model for each fixed value of $\tau$ and $A$.
Uncertainty estimates for $\tau$ and $A$ represent 68\% confidence, based on a change of $\Delta\Sigma(\chi^2)$=1. The resulting light curves from this model are shown in Figure~\ref{fig:UVW2linklightcurves}.

The UVW2-linked soft excess model shows a significant improvement in $\chi^2$ versus the fixed-$\Gamma$ model in Section \ref{sec:fixedgamma}, indicating that the shape of the power law continuum is intrinsically variable and should be a free parameter. The ICCF results in Figure~\ref{fig:UVW2linklightcurves} and Table~\ref{table:peaklags} show lags consistent with zero between the UVW2 band and both the total 0.3--8 keV NICER X-ray flux ($R_{\mathrm{Peak}}=0.60$) and the flux of the X-ray power law component ($R_{\mathrm{Peak}}=0.50$). The power law index and UVW2 flux are weakly correlated with $R_{\mathrm{Peak}}=0.54$ at $\tau=-3.2\substack{+1.0\\ -2.2}$ days, discussed in Section~\ref{sec:gammadiscuss}. 

Given that the UVW2-linked soft excess model has stronger physical motivation than the fixed-$\Gamma$ model and provides a better fit to the data, we proceed with the light curves presented in Figure~\ref{fig:UVW2linklightcurves} and Table \ref{table:uvlinkparam} for our subsequent analysis.

Out of the 383 epochs, 19 have a $\chi^2_{\nu}>2$ (e.g. MJD 58272 in Table~\ref{table:uvlinkparam}). These spectra have large residuals from $\sim12$--15 keV, where the observed count rates exceed the power law prediction of the SCORPEON model's non-X-ray background component (see Figure~\ref{fig:spec_comp}). Since the effective area of NICER drops precipitously above 12 keV \citep{remillard_2022}, these residuals indicate a transient component of the high-energy background that is not yet included in the SCORPEON model. The fit improves in the 0.3--10 keV band, including from 5--10 keV where the background is typically brighter than the AGN. This suggests that the background component is only present above 12 keV and has a minimal impact on our models of the AGN source.

\begin{deluxetable*}{ccccccccc}
\tablewidth{0pt}
\tablecaption{\label{table:uvlinkparam}Spectral Analysis Results: NICER background-subtracted count rate, hardness and spectral parameters for the UVW2-linked soft excess model in Section~\ref{sec:tiedsoftexcess}. Fluxes are calculated from 0.3--8 keV. The first five observations are listed, with uncertainties corresponding to the 68\% confidence interval ($1\sigma$) for each parameter. The full table is published in machine-readable format.}
\tablehead{
\colhead{Obs. Date} &
\colhead{Count Rate} &
\colhead{Hardness Ratio} &
\colhead{Total Source} &
\colhead{Power Law} &
\colhead{Power Law} &
\colhead{Soft Excess} &
\colhead{Best fit} & 
\colhead{Degrees}\\ 
\colhead{(MJD)} &
\colhead{(counts~s$^{-1}$)} &
\colhead{H=1.5--8 keV}&
\colhead{$\log(\Phi_{\mathrm{PL}}+\Phi_{\mathrm{SE}}+\Phi_{\mathrm{K}\alpha})$} &
\colhead{$\log(\Phi_{\mathrm{PL}})$} &
\colhead{Index $\Gamma$} &
\colhead{$\log(\Phi_{\mathrm{SE}})$} &
\colhead{$\chi^2_{\nu}$} &
\colhead{of}\\ 
\colhead{} &
\colhead{0.3-8 keV} &
\colhead{S=0.3--1.5 keV} &
\colhead{(erg~cm$^{-2}$~s$^{-1}$)} &
\colhead{(erg~cm$^{-2}$~s$^{-1}$)} &
\colhead{} &
\colhead{(erg~cm$^{-2}$~s$^{-1}$)} &
\colhead{} &
\colhead{Freedom}
}
\startdata
58256.95 & $32.10\pm0.17$ & $-0.641\pm0.006$ & $-10.233\pm0.003$ & $-10.248\pm0.003$ & $2.10\pm0.01$ & $-11.739\pm0.008$ & $115.28$ & $126$ \\ 
58257.00 & $31.14\pm0.22$ & $-0.652\pm0.009$ & $-10.247\substack{+0.006\\ -0.004}$ & $-10.263\substack{+0.006\\ -0.004}$ & $2.13\substack{+0.01\\ -0.02}$ & $-11.739\pm0.008$ & $111.59$ & $116$ \\ 
58272.38 & $37.25\pm0.20$ & $-0.660\pm0.006$ & $-10.203\substack{+0.002\\ -0.003}$ & $-10.222\substack{+0.002\\ -0.003}$ & $2.16\pm0.01$ & $-11.619\pm0.008$ & $392.01$ & $158$ \\ 
58293.62 & $31.44\pm0.20$ & $-0.663\pm0.008$ & $-10.278\pm0.004$ & $-10.296\pm0.004$ & $2.17\pm0.01$ & $-11.738\pm0.008$ & $166.26$ & $157$ \\ 
58294.65 & $31.56\pm0.18$ & $-0.649\pm0.007$ & $-10.267\pm0.003$ & $-10.283\pm0.003$ & $2.13\pm0.01$ & $-11.760\pm0.008$ & $132.15$ & $98$ \\ 
... & ... & ... & ... & ... & ... & ... & ... & ...
\enddata
\end{deluxetable*}
\subsection{Variability Amplitudes and Timescales}
\label{sec:fvaranalysis}
We quantify the variability in each light curve from Section~\ref{sec:tiedsoftexcess} using the mean fractional variation $F_{\mathrm{var}}$ (\citealt{Rodriguez_1997}, \citealt{vaughan03}):

\begin{equation}
    F_\mathrm{var}=\sqrt{\frac{S^2-\overline{\sigma^2_{\mathrm{err}}}}{\overline{\Phi}^2}}
\end{equation}
\noindent where $\overline{\Phi}$ is the mean flux (or $\overline{\phi}$ for the mean UVW2 flux density), $\overline{\sigma^2_{\mathrm{err}}}$ is the mean square error, and $S^2$ is the sample variance. Across the full campaign and without detrending, $F_{\mathrm{var}}$ values for the NICER total X-rays, the power law, the soft excess, and the Swift UVW2 band are 19\%, 18\%, 49\%, and 25\%, respectively.  

We test if the UVW2 variability can be attributed to X-ray reprocessing, taking the relative energies of each band into account (see \citealt{Uttley_2003}). We use $F_\mathrm{var}$ of the 0.3--8 keV power law and UVW2 fluxes as  proxies for the variability in the full X-ray power law and UV continua, respectively. Spectral Energy Distribution (SED) modeling of Fairall~9 in \citealt{hagendone23} showed that ~77\% of the total accretion power is emitted by the disk in the UV. The X-rays, whose dominant contribution comes from the power law, only contribute ~23\%. If the UV component contributes 77\% of the total flux and 25\% of this power is variable, then the variable component of the X-rays must contribute at least 19\% (i.e., 77\%$\times$25\%) of the total energy of the AGN to drive the UV light curve through reprocessing alone. However, the X-rays contribute 23\% of the total energy but only 18\% of this is variable, meaning that only $\sim4$\% of the total accretion power comes from the variable X-ray component. This suggests that the hot X-ray corona is not energetic or variable enough to to drive the fractional variability amplitude seen in the UV.

\begin{figure*}[t]
  \centering
  \includegraphics[width=\textwidth]{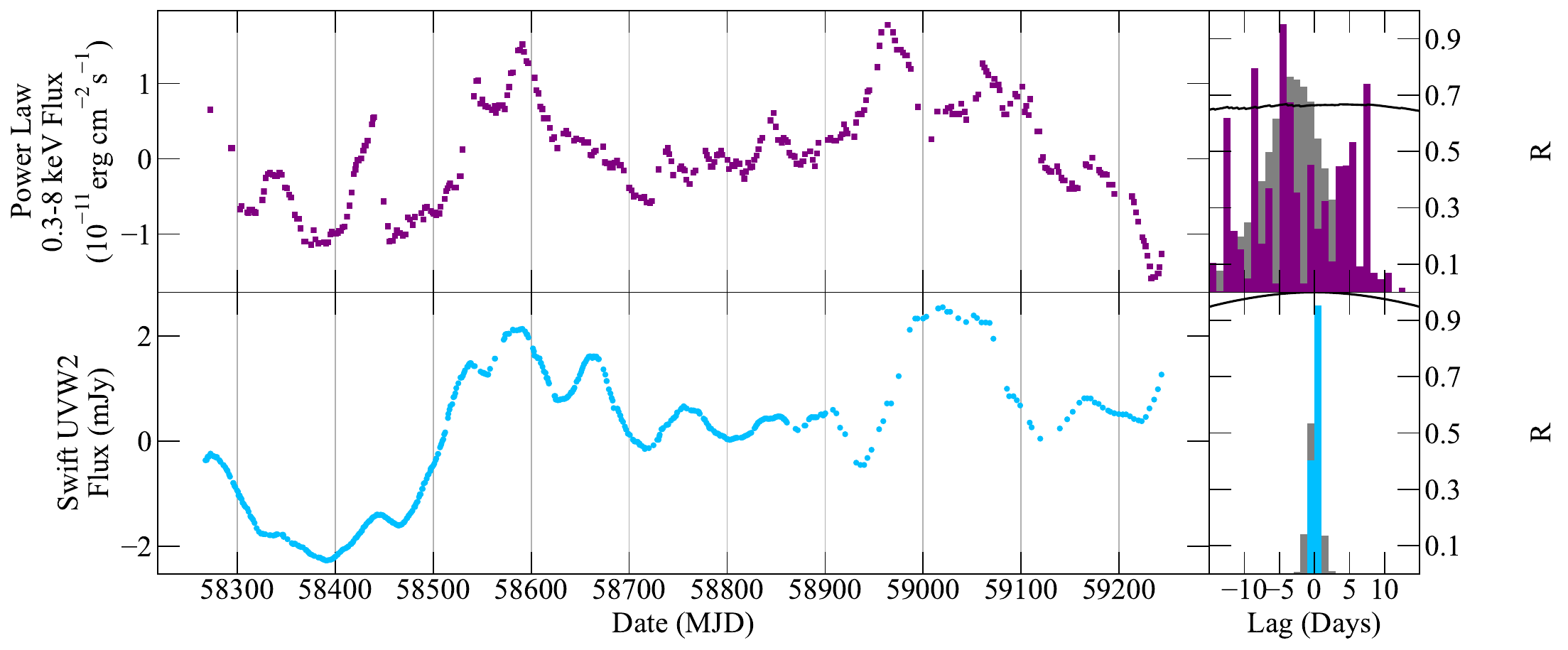}
  \caption{Left: Light curves showing slow variability in the X-ray power law flux (top, purple) and UVW2 flux density (bottom, blue), adapted from Fig.~\ref{fig:fixedgammalightcurves} and smoothed using a rolling boxcar average at each epoch with a width of 30 days. Right: Cross-correlation functions calculated using UVW2 as the reference band, with histograms representing the peak (colored) and centroid (gray) probability distributions. The peak X-ray/UV correlation is $R_\mathrm{Peak}=0.66$ at a lag of $\tau=-3.4 \substack{+8.8\\ -6.0}$ days.}
  \label{fig:30daysmoothslow}
\end{figure*}

\begin{figure*}[t]
  \centering
  \includegraphics[width=\textwidth]{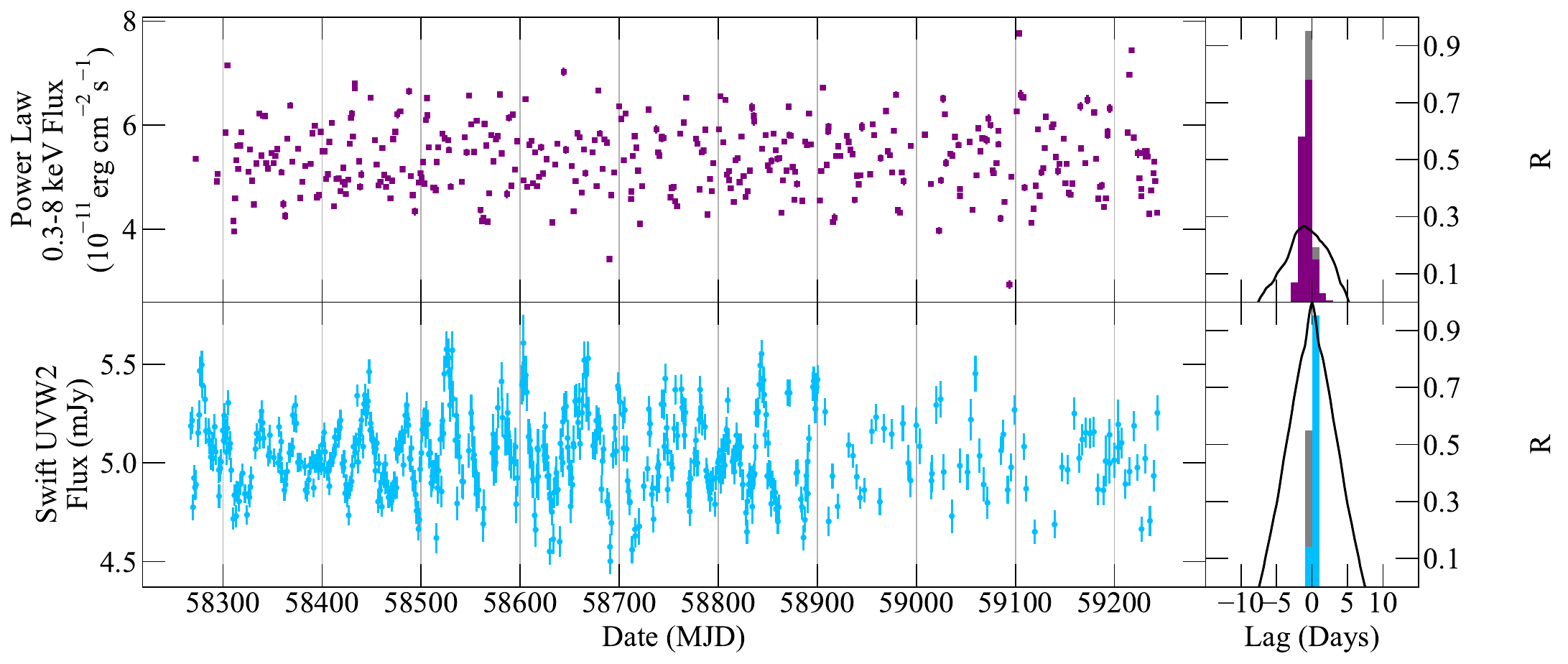}
  \caption{Same as Fig.~\ref{fig:30daysmoothslow} but for the fast variability component of the light curves, calculated by subtracting a 30-day-wide boxcar average from the original light curves in Fig.~\ref{fig:fixedgammalightcurves}. The X-ray/UV correlation is not statistically significant, peaking at $R_\mathrm{Peak}=0.27$ with a lag of $\tau=-1.0\pm 0.8$ days.}
  \label{fig:30daysmoothfast}
\end{figure*}

Analysis of the first year of UV and optical variability in the campaign revealed trends that last hundreds of days without a corresponding X-ray component \citep{Hernandez2020} and optical-leading-UV lags of $\sim70$ days \citep{Yao_2023}. This motivates our search for sources other than X-ray reprocessing that may contribute to the UV variability. We isolate features in both the X-ray and UV light curves on multiple timescales by separating ``fast" and ``slow" trends in the data. The slow component is calculated from the observed light curve, standardized to a mean of zero. We take a rolling boxcar average with full width $t$ days at each epoch, smoothing out variability on timescales shorter than $t$ days (Figure~\ref{fig:30daysmoothslow}). The isolated fast component is the residual between the observed light curve and the slow component (Figure~\ref{fig:30daysmoothfast}). We measure the X-ray/UV  ICCF and $F_{\mathrm{var}}$ for each pair of fast light curves using a common smoothing width $t$, testing a range of $t=2$--200 days. To ensure an equal boxcar width across the sample, we exclude epochs within $t/2$ days of the first and last dates of the campaign.

We compare the energetic contributions of the fast X-ray power law and UV components, again using the percentages of the total AGN power calculated from SED modeling in \cite{hagendone23}. For example, by smoothing over $t=10$ days, the total X-ray power law and UVW2 light curves have $F_{\mathrm{var}}=$~18.5\% and 25.2\%, respectively. The fast components have $F_{\mathrm{var}}=$~8.9\% (X-ray) and 1.5\% (UV), corresponding to 2.0\% and 1.2\% of the total AGN power. 

\begin{figure}[t]
  \centering
  \includegraphics[width=\columnwidth]{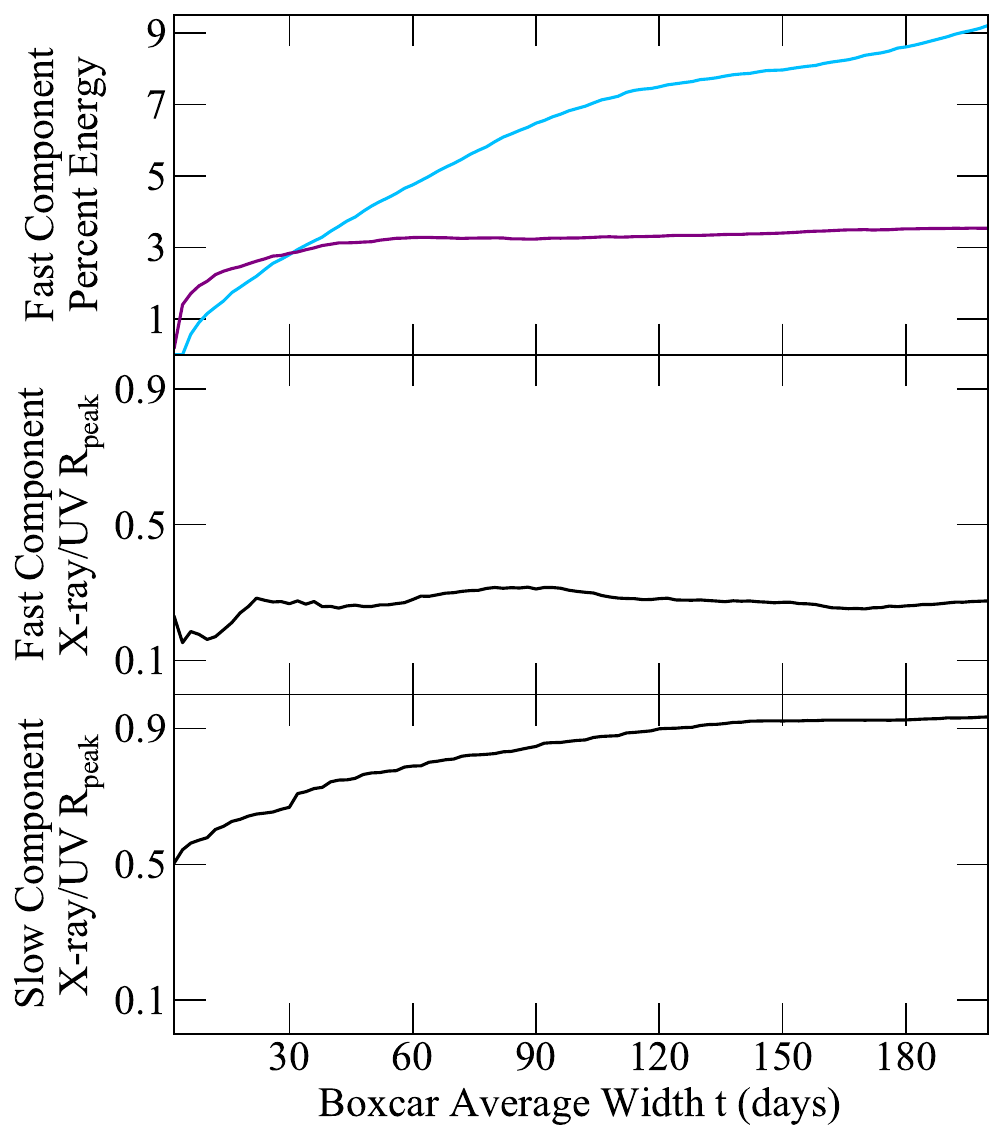}
  \caption{Top: Energy contributions from the fast variability components of the X-ray (purple) and UVW2 (blue) light curves in Figure~\ref{fig:UVW2linklightcurves}, which are detrended using a rolling boxcar average with a width of $t=2$--200 days for each epoch. Energies are calculated as a percentage of the total AGN power based on $F_{\mathrm{var}}$ (Equation~2) and the SED modeling results of \cite{hagendone23}. Middle: $R_\mathrm{Peak}$ between the fast variability components of the X-ray and UV light curves after detrending with a common boxcar width $t$. Bottom: $R_\mathrm{Peak}$ between the slow variability components of the X-ray and UV light curves, produced using a boxcar average with full width $t$.}
  \label{fig:fvarfastslowr}
\end{figure}

The energy of the fast X-ray component exceeds that of the fast UV on smoothing timescales shorter than 30 days, as shown in Figure~\ref{fig:fvarfastslowr}. For $t=30$ days, the fast X-ray component has $F_{\mathrm{var}}=$~12.3\% and contributes 2.8\% of the total AGN power. This equals the energy of the UVW2 component, which has $F_{\mathrm{var}}=$~3.6\%. This suggests that the X-rays can power the fast UV disk variability on timescales of days to weeks, while slow trends in the UV on timescales of several months to years require an alternative explanation (see Section~\ref{sec:fvardiscuss}.)

We compare the relative flux contributions of the X-rays and UV during our campaign to the SED modeled in Figure~6 of \cite{hagendone23}, which shows that the X-ray power law predominantly emits from 0.1--100 keV. We extrapolate our mean 0.3--8 keV flux ($\Phi_{\mathrm{PL}}=5.35 \times 10^{-11}$ erg~cm$^{-2}$~s$^{-1}$) using the mean power law slope. Accounting for the standard deviation in $\Gamma$ of $\pm 0.08$ across the campaign, we estimate an uncertainty in $\Phi_{\mathrm{PL}}$ of $\substack{+3\% \\-1\%}$. Including a cutoff energy ($E_\mathrm{cut}$) in the power law model also decreases the extrapolated flux. Adopting the value of $E_\mathrm{cut}=100$ keV used in the SED modeling of Fairall~9 by \cite{hagendone23} decreases the extrapolated $\Phi_{\mathrm{PL}}$ by 6\%. The higher
estimate of $E_\mathrm{cut}=784\pm \substack{+162\\ -271}$ obtained by \cite{Lohfink_2016} via joint modeling of an XMM+NuSTAR spectrum decreases $\Phi_{\mathrm{PL}}$ by $<1\%$. 

We combine the uncertainties estimated by varying $\Gamma$ and setting $E_\mathrm{cut}=100$ to yield the mean 0.1--100 keV flux of $\Phi_{\mathrm{PL}}=1.04 \pm \substack{+0.03\\ -0.07} \times 10^{-10}$ erg~cm$^{-2}$~s$^{-1}$. Smoothing by $t=30$ days results in a detrended fast power law component with $F_\mathrm{var}=$~12.3\% and a corresponding flux contribution from 0.1--100 keV of $1.28 \pm \substack{+0.04\\ -0.08} \times 10^{-11}$ erg~cm$^{-2}$~s$^{-1}$. The power law spectrum is dominated by the soft X-rays, with $78\%\pm 6 \%$ of the flux originating from the 0.1--10 keV band.

The mean flux density $\phi_\nu$ of the Swift UVW2 band is $5.042 \pm 0.003$ mJy. For the central wavelength of 1928 \AA{} ($\nu=1.55 \times 10^{-15}$ Hz), we convert from $\phi_\nu$ to an extinction-corrected flux of $\nu\phi_\nu = \Phi_\mathrm{1928}=9.269\pm0.005 \times 10^{-11}$ erg~cm$^{-2}$~s$^{-1}$ using the extinction relationship of \cite{Cardelli_1989} and $E(B-V)=0.022$ \citep{Schlafly11}. Using a boxcar detrending width of 30 days, the fast variability component of the UVW2 light curve has an amplitude of $F_\mathrm{var}=$~3.6\%. Its corresponding flux contribution is $3.34 \times 10^{-12}$ erg~cm$^{-2}$~s$^{-1}$. This is much less than the variable X-ray flux, indicating that X-ray reprocessing can plausibly power the fast UVW2 variability and leave a large energy residual to drive variability at longer wavelengths. 

\begin{figure*}[t]
  \centering
  \includegraphics[width=\textwidth]{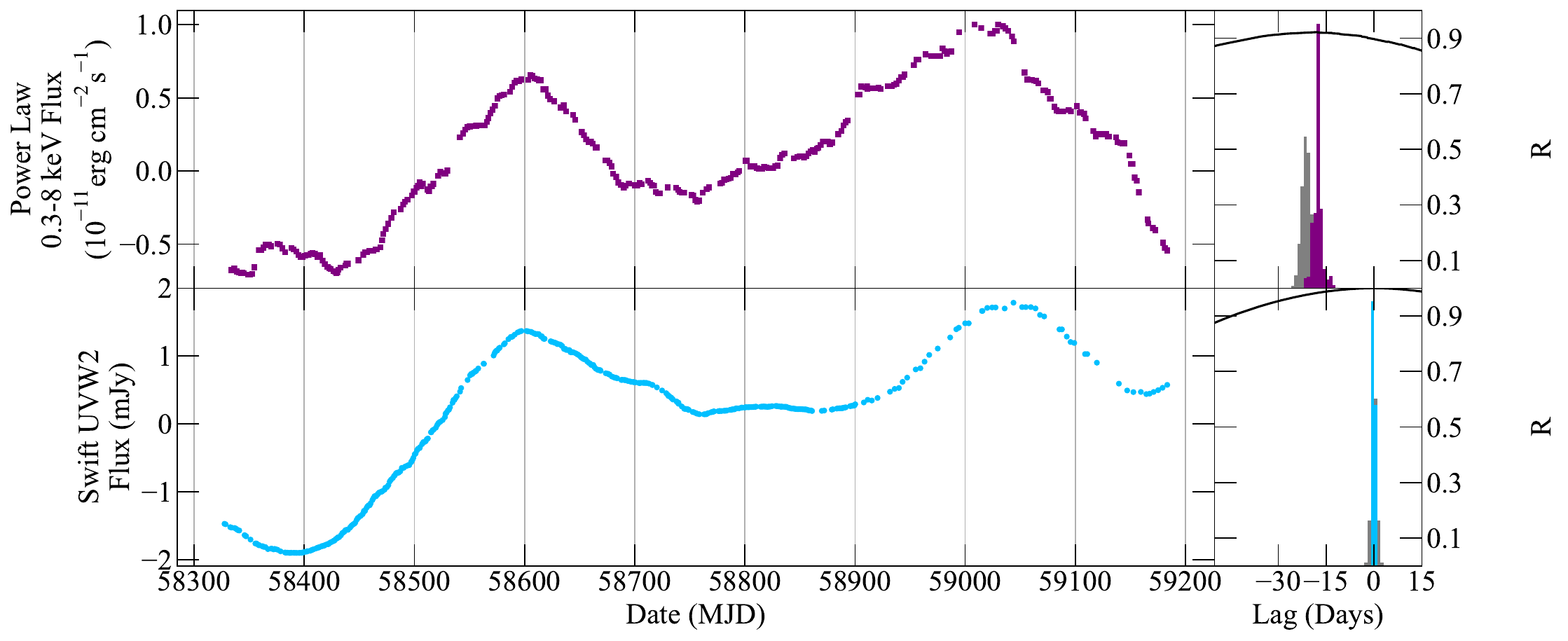}
  \caption{Same as Fig.~\ref{fig:30daysmoothslow}, but for slow variability isolated with a boxcar width of 150 days. The X-ray/UV correlation is very high ($R_\mathrm{Peak}=0.92$) and peaks at a lag of $\tau=-17.8 \substack{+1.4\\ -1.0}$ days, indicating that the X-rays lead the UV on long timescales. The ICCFs are extremely flat as a consequence of the broad boxcar width.}
  \label{fig:150daysmoothslow}
\end{figure*}

We also measure the X-ray/UV correlation between multiwavelength pairs of fast and slow light curves for each smoothing timescale, presented in Figure~\ref{fig:fvarfastslowr}. The fast light curves are always weakly correlated, with $R_\mathrm{Peak}$ values between 0.2--0.3 and lags consistent with zero. However, the correlation between the slow X-ray and UV light curves increases steeply at larger smoothing timescales, reaching a plateau of $R_\mathrm{Peak}=0.92$ for $t>150$ days. The negative lag measurement of $\tau_\mathrm{peak}=-17.8 \substack{+1.4\\ -0.4}$ days ($\tau_\mathrm{cent}=-21.7 \substack{+1.4\\ -1.5}$ days) between the slow light curves smoothed over $t=150$ days (Figure~\ref{fig:150daysmoothslow}) strongly suggests that long-term trends in the X-rays lead the UV. This is opposite to the direction expected for inbound propagation \citep{Arevalo_2006}. The slow component of the X-rays also lacks sufficient energy to directly power these UV changes at $t=150$ days, since it contributes only 0.6~\% of the total AGN power versus 11.7~\% in the UV. The implications of this are discussed in Section~\ref{sec:fvardiscuss}.

\section{Discussion}
\label{sec:Discussion}
\subsection{Multi-timescale Variability in Fairall~9}
\label{sec:fvardiscuss}
By isolating smooth features in the X-ray and UV light curves of Fairall~9 we reveal behavior on timescales of days to months, with an increasing inter-band correlation as the light curves are averaged over longer timescales (Figure~\ref{fig:fvarfastslowr}). Our analysis of detrended light curves in Section~\ref{sec:fvaranalysis} suggests that the X-ray flux of the power law component, attributed to the hot corona, is sufficient to power fast variability on timescales shorter than $\sim$30 days in the UVW2 continuum, attributed to the accretion disk. This is consistent with the standard X-ray reprocessing scenario. However, correlations between the X-ray/UV are much weaker than expected ($R\simeq0.3$) for a static, isotropic X-ray emitter. This may be caused by dynamic variability in the height, particle density, or energy density of the X-ray corona, that can weaken the observed X-ray/UV correlation while still producing strongly correlated UV/optical light curves \citep{Panagiotou_2022_xraycorrel}.

We also observe variations on timescales of 30--150 days in the X-rays which are not matched by the UVW2, such as the prolonged bright state from MJD 58400--58480 and the double peaked feature from MJD 58900--59100 (Figure~\ref{fig:30daysmoothslow}). This suggests an additional process in the corona not linked to X-ray reprocessing, such as a possible structural change, which may contribute to the weak correlation between fast X-ray and UV variability. The correlation would also be reduced if an optically thick layer lies between the hot corona and the disk, which would act as an intermediate reprocessor and distort the light curve incident on the disk (see Section~\ref{sec:geomdiscuss} for further discussion). A similar explanation was invoked in the case of Mrk 817, in which the X-ray/UV correlation was insignificant during an epoch of heightened absorption \citep{partington23}.

Our analysis demonstrates that the X-rays and UV undergo similar trends on smoothing timescales of 150--200 days, with $R_\mathrm{Peak}\sim0.92$ (Figure~\ref{fig:fvarfastslowr}). Although the X-rays lead the UV by $\sim17$ days, the energy of the X-rays is insufficient to power the high amplitude of variability seen in the UVW2 light curve through reverberation at this timescale. Thus we must consider alternative explanations for the origin of this correlated variability.

A parabolic trend in the UV and optical light curves is present for the first year of the campaign, as discussed in \cite{Hernandez2020}. This is also visible in the smoothed UV light curves presented in Figures~\ref{fig:30daysmoothslow} and \ref{fig:150daysmoothslow}. Using the first year of data covering the Swift UVW2 to optical $z$ bands, \cite{Yao_2023} found negative lags of $\sim70$ days, in which the optical leads the UV, plus a short lag ($<10$ days) in the UV leading optical direction consistent with the reverberation picture. Analysis of the same dataset by \cite{Neustadt_2022} suggests that variability on timescales of $\sim200$ days may be caused by temperature perturbations in the disk that travel radially inward and outward, producing the observed long-term trend in the UV and optical light curve. The duration of these features and their apparent lag suggest that there exist intrinsic fluctuations in the disk which travel at speeds much slower than $c$.

A bright feature in the X-ray light curve that starts near MJD 58400 and lasts for nearly 80 days (Figure~\ref{fig:30daysmoothslow}) complicated the analysis of the X-ray/UV connection on long timescales during the first year of monitoring. However, in the full campaign it is apparent that the smoothed X-ray light curve shares many features with the slow variability in the UV. A positive lag, in which the UV leads the X-rays, is expected for the scenario in which slow, inbound mass fluctuations reach the inner disk and modulate the availability of seed photons for the X-ray corona (e.g., \citealt{Arevalo_2005}, \citeyear{Arevalo_2008}). Our negative lag measurement on timescales of 150 days is in the opposite direction of this expectation. However, the  ICCF is very broad as a consequence of the boxcar average method, so the lag measurement may not be reliable. In Figure~\ref{fig:150daysmoothslow}), it is also apparent by eye that the UV leads the X-rays during the ascent to the first peak in flux near MJD 58600, while the X-rays lead the UV during the second peak near MJD 59000. As such, caution should be used when interpreting this lag since it may not be physically meaningful. 

Our reverberation study demonstrates that the X-rays can plausibly drive the UVW2 fluctuations in Fairall 9 on timescales of days to months, while variability in the disk is required to explain changes in the light curve on longer timescales. We note that similar cases of variability with a fast reverberation component and an additional slow trend are also found in other AGN, e.g. NGC~5548 (\citealt{Uttley_2003}, \citealt{Panagiotou_2022_5548}). Further analysis of Fairall~9 using the light curves from this campaign, including models of the multiwavelength power spectra and the variable SED, would improve upon our estimates of the separate energetic contributions made by X-ray reprocessing and intrinsic disk variability.

\subsection{The Nature of the Soft Excess}
\label{sec:wcdiscuss}
In Section~\ref{sec:fixedgamma}, we found a much stronger correlation between the X-ray soft excess and UVW2 light curves ($R=0.82$) than the X-ray power law and UVW2 ($R=51$). This revealed for the first time that the variability in the soft excess shares a similar amplitude and pattern with the UV accretion disk. Common fast variability features measured in Section~\ref{sec:tiedsoftexcess} suggest the presence of a reverberation signal, with the soft excess leading the UV. Both light curves also demonstrate similar variability on timescales of months, with crests and troughs between MJD 58400--58600 and MJD 58950--59150 that have a different shape than those seen in the X-ray power law light curve.

A possible explanation for the behavior of the soft excess is that its predominant emission source is linked directly to the accretion disk, interior to the region of peak UVW2 emission. This area would experience the same slow temperature fluctuations seen in the UV disk. It would also respond to the central fast reverberation signal before the UVW2 continuum, consistent with the measured negative lag. In this interpretation, a significant portion of the soft excess emission could in fact be the high-energy spectral tail of the upper layers of the disk (a so-called ``warm corona," see Fig.~\ref{fig:schematic}). 

The observed soft excess/UV correlation is consistent with the model of a passive inner disk sandwiched between optically thick layers of Comptonizing material, as presented by \cite{Petrucci_2018} and expanded on for Fairall~9 in \cite{hagendone23}. This portion of the disk would serve as a source of seed photons which scatter off of the warm corona, producing primarily EUV and some soft X-rays. If variability in the X-ray soft excess originates from a warm corona with emission extending into the EUV, the total power may be much higher than the X-rays, surpassing even the UV disk. Detailed SED modeling would also be necessary to separate the EUV contributions of the hot and warm coronae.

Any estimate on the radiative power of the warm corona from the flux of the soft excess would be an upper limit, however, as the analysis of XMM+NuSTAR spectra by \cite{Lohfink_2016} shows the presence of blurred reflection features from an ionized inner disk which contribute flux at soft energies. This is in addition to the warm Comptonization component in the soft X-ray spectrum which we interpret as the warm corona. This reflection also implies that the disk extends near the innermost stable circular orbit rather than becoming truncated, and that the warm corona must somehow become optically thin, possibly as a continuously evolving layer which transitions into an optically thin hot corona at small radii. 

Using the joint XMM-NuSTAR observation from 2014, \cite{Lohfink_2016} attribute 55\% of the flux at 1 keV to warm Comptonization and 10\% to blurred ionized reflection. This suggests that the warm corona is the dominant emission source in the soft excess feature. The remaining 35\% of the flux is attributed to the power law. In our analysis with NICER, however, the power law component is always brighter than the soft excess (see Fig.~\ref{fig:spec_comp}), in direct contrast to \cite{Lohfink_2016}. Since the soft excess is highly variable, it is possible that the 2014 observation captured the warm corona in a much brighter state than during our campaign from 2018--2021. Alternatively, this disagreement may reflect bias introduced by model used for the soft excess, given the degeneracy between the soft excess flux and the power law slope. 

Greater insight into the relative contribution of reflection in the soft excess over long timescales could be obtained via a more complex spectral model, which would incorporate both ionized reflection and a warm corona. We are restricted to a simple model in this campaign due to the quality of our spectra (see Section~\ref{sec:spectralmodel}), but this goal could be achieved with a more sensitive instrument (e.g., the proposed STROBE-X, \citealt{ray_2019}) or longer exposure times with NICER. 

It is also likely that the warm corona is not present in every AGN, as \cite{Done12} predicts that the soft excess due to Comptonization becomes more significant with increasing Eddington ratio $\dot{m}_\mathrm{Edd}=L_\mathrm{Bol}/L_\mathrm{Edd}$. The evidence for a warm corona in Fairall~9 contradicts the prediction that this region forms only at high $\dot{m}$, given its relatively low $\dot{m}_\mathrm{Edd}= 0.06$ \citep{hagendone23}. Future reverberation mapping experiments with NICER  on AGN across a range of $\dot{m}_\mathrm{Edd}$ are needed to observationally determine the conditions under which the warm corona can form, including lower limits on $\dot{m}_\mathrm{Edd}$. It is also possible that observing a higher-$\dot{m}_\mathrm{Edd}$ source would improve the measurement of the soft excess flux with NICER, potentially even overcoming the degeneracy between $\Phi_\mathrm{SE}$ and $\Gamma$ described in Section~\ref{sec:spectralmodel}.

\subsection{Geometry of the Inner Accretion Flow}
\label{sec:geomdiscuss}
\begin{figure}[t]
  \centering
  \includegraphics[trim={5cm 4.25cm 11.25cm 9cm},clip,width=\columnwidth]{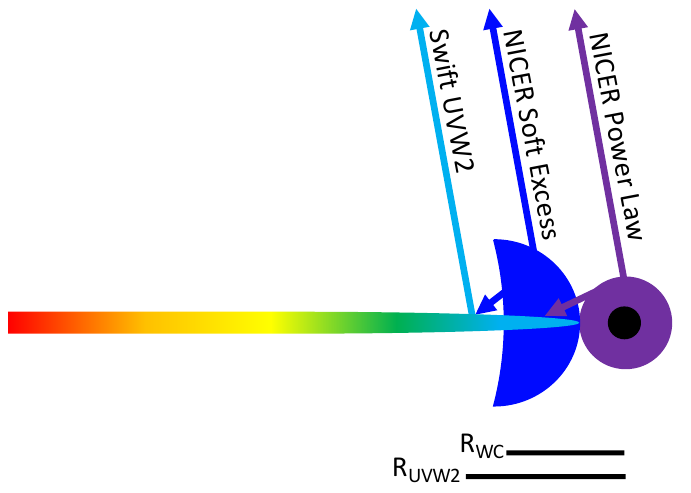}
  \caption{Schematic of the accretion flow of Fairall~9. The thermal accretion disk is shown in a gradient from red to light blue. The peak UVW2 disk emission in the direction of the observer is represented by a blue arrow originating at $R_\mathrm{UVW2}$. The warm corona (dark blue) layers the thermal disk, extending to $R_\mathrm{WC}$. A dark blue arrow represents emission from the warm corona, plus X-rays from the hot corona (purple) which are gravitationally redshifted and reflected off of the inner disk as soft X-rays.}
  \label{fig:schematic}
\end{figure}

We recover a lag between variability in the X-ray soft excess and UVW2 continuum of $\tau=-1.2 \substack{+0.3\\ -0.1}$ days from our grid search in Section~\ref{sec:tiedsoftexcess}. Since variability in the UV/optical disk of Fairall~9 is shown to propagate outwards to radii emitting at longer wavelengths on timescales of a few days \citep{Hernandez2020}, this would place the warm corona emission interior to the UVW2 disk, at $R_\mathrm{WC}$ (see Figure \ref{fig:schematic}). If this lag is equivalent to the light travel time between the two regions of peak emission in each wavelength band, this can be converted into a distance in gravitational radii of $R_\mathrm{UVW2}-R_\mathrm{WC}=81 \substack{+20\\ -7} R_\mathrm{G}$ using $M_\textit{BH}=2.6 \times 10^8 M_\odot$ \citep{peterson04}. 

We can also predict the theoretical location of the UVW2 disk using Equation (12) of \cite{fausnaugh16}, which relates the expected distance $R$ from the central BH to the peak wavelength $\lambda$ of thermal emission. The function is parameterized by the radiative efficiency of accretion $\eta$, the ratio of internal viscous heating to external radiative heating $\kappa$, and a factor $X$ which relates the wavelength to temperature at $R$, considering that a large range of radii will emit some flux with the same $\lambda$: 

\begin{equation} 
    R=\left(X\frac{k\lambda}{hc}\right)^{4/3}\left[\left(\frac{GM}{8\pi\sigma}\right) \left(\frac{L_\mathrm{Edd}}{\eta c^2}\right)(3+\kappa)\dot{m}_\mathrm{Edd}\right]^{1/3}. 
\end{equation}
We assume $\kappa=0$ due to negligible X-ray heating and $\eta=0.1$, as is typical in analysis of other AGN (e.g. \citealt{edelson2017}). We set $\dot{m}_\mathrm{Edd}= 0.06$ \citep{hagendone23}. We compare values of $X$ calculated using the Planck function ($X=2.49$, \citealt{fausnaugh16}) and taking into account time-variable temperature fluctuations ($X=5.04$, \citealt{tie_2018}). For the central $\lambda=1928$ \AA{}, $R_\mathrm{UVW2}=32 R_\mathrm{G}$ (0.34 light days) if $X=2.49$ or $R_\mathrm{UVW2}=84 R_\mathrm{G}$ (0.86 light days) if $X=5.04$. We would then expect the warm corona to be located above the inner disk, interior to $R_\mathrm{UVW2}$ with a height of tens of $R_\mathrm{G}$. Depending on the vertical extent of this optically thick layer, the hard X-rays may also be blocked from reaching the disk, causing the warm corona to act as an intermediate reprocessor. 

\cite{petrucci13} (hereafter P13) also found a warm corona to be responsible for the soft excess of Mrk~509, invoking a slab-like geometry in the outer layer of the accretion disk. Similar to the proposed geometry of Fairall~9 in Figure~\ref{fig:schematic}, P13 suggests that the accretion flow is continuous and transitions from the warm to the hot corona phase at $\sim 10 R_\mathrm{G}$. In the slab geometry of P13, the warm corona covers a large area of the disk, while our prediction for Fairall~9 suggests that it lies within $\sim 80 R_\mathrm{G}$ of the SMBH. Our lag measurements indicative of a vertically extended corona also contrast with the geometry of Mrk~509 in P13, which places the vertical edge of the warm corona at a lower height than the hot corona. Our analysis is consistent with the finding of P13 that although the hot corona irradiates the warm corona, the primary heating source of the warm corona is internal.

\subsection{Power Law Spectral Variability}
\label{sec:gammadiscuss}
The UVW2 flux density and the spectral index of the power law model component are also weakly correlated ($R_{\mathrm{Peak}}=0.54$) at $\tau=-3.2\substack{+1.0\\ -2.2}$ days. We expect that higher UVW2 fluxes will contribute more seed photons to the hot corona, cooling the electrons through increased Comptonizaton and producing the ``softer-when-brighter" power law behavior common to AGN coronae \citep{Magdziarz_1998}. However, the negative lag indicates that the power law shape changes before the UVW2 (though the lag is not highly significant), so the influx of seed photons cannot be the dominant cause of variability. 

It is possible that an additional variability process occurs in the soft excess on timescales of a few days and is not characterized by the linear scaling relationship in Equation \ref{eq:SEUVfluxscal}. Due to the degeneracy between $\Phi_\mathrm{SE}$ and $\Gamma$, this may then influence the shape of the modeled power law. Alternatively, if the correlated variability between $\phi_{\nu,\mathrm{UVW2}}$ and $\Gamma$ is intrinsic to the hot X-ray corona, this may indicate that the production of soft X-ray photons is the key indicator of the X-ray reprocessing scenario, due to these photons being more readily absorbed by the disk. However, as described in Section \ref{sec:geomdiscuss}, a lag of three days greatly exceeds the expected light travel time between the UVW2 disk and the hot corona. In this light, we favor the former scenario in which the variability of $\Gamma$ serves as a proxy for some slow process in the warm corona above the inner disk, which occurs on timescales of a few days in addition to the linear process described by Equation \ref{eq:SEUVfluxscal}.

\section{Summary and Conclusions}
\label{sec:Conclusion}
Our near-daily monitoring of Fairall~9 for over three years uses spectral analysis of individual 1 ks NICER X-ray observations in combination with Swift UVW2 photometry to reveal that:

\begin{itemize}
    \item The X-rays are variable and energetic enough to drive variability in the UVW2 on timescales shorter than 30 days. This suggests that X-ray reprocessing is the primary cause of fast variability in the UVW2, despite a weak X-ray/UV correlation ($R\simeq0.3$). Energetic estimates are based on the SED modeling of Fairall~9 in \cite{hagendone23}.
    \item A pattern of coherent UV and optical variability on timescales longer than 150 days which progresses inwards, opposite to the reverberation signal, (\citealt{Hernandez2020}, \citealt{Yao_2023}), is likely connected to the correlated slow variability observed in the X-ray and UV light curves ($R\simeq0.9$) across the full campaign. The X-ray corona lacks the requisite variability at these timescales to drive the UV through reprocessing. This suggests an origin within the disk (e.g. intrinsic temperature perturbations, \citealt{Neustadt_2022}) which drives the slow X-ray trends through the production of seed photons.
    \item Variability on the order of days observed in the Swift UVW2 band is more closely related to changes in the X-ray soft excess than the X-ray power law continuum. Lags between the soft excess and the UVW2 continuum of $\tau\simeq1.2$ days suggest a scenario in which the soft X-rays are produced by a layer of Comptonizing electrons above the inner accretion disk (i.e. the warm corona). This region likely emits into the observable EUV, with the soft excess manifesting as the high-energy spectral tail. 
    \item While some contribution to the soft excess from blurred reflection is certainly present \citep{Lohfink_2016}, this is not the component which dominates variability on the timescales of our experiment.
\end{itemize}

Future studies on AGN with a stronger soft excess (expected from higher $\dot{m}_\mathrm{Edd}$ sources) should aim to overcome the degeneracy between the flux of the soft excess and the shape of the power law spectrum. In Fairall~9, this led to ambiguous results on the nature of variability in the soft X-rays which led the UVW2 light curve on timescales of a few days, longer than the expected light travel delay between the regions. Overall, this study verified the effectiveness of NICER for a new type of reverberation mapping experiment on the separate X-ray emission components of AGN, thanks to the instrument's unprecedented combination of flexible scheduling capabilities and high throughput and large effective area. 

\begin{acknowledgments}
ERP and EC gratefully acknowledge support for this work from NASA grant number 80NSSC21K0635. We also thank Scott Hagen for his helpful discussion.
\end{acknowledgments}
\facilities{NICER, Swift}
\software{Astropy \citep{Astropy_2013},
          Matplotlib \citep{Hunter_2007}, SciPy \citep{Scipy_2020}, PyCCF (\citealt{Peterson_1998}, \citealt{Sun_2018}), HEASOFT \citep{Blackburn_1995}, ftgrouppha \citep{kaastra16}, NICER 3C50 Estimator \citep{remillard_2022}, Swift XRT Light Curve Generator (\citealt{evans07}, \citealt{evans09}),  XSPEC \citep{arnaud96}, \textsc{tbabs} \citep{Wilms_2000}}

\newpage
\appendix
\twocolumngrid
\section{The NICER 3C50 Background Estimator and SCORPEON Background Model}
\label{sec:bgappendix}
Since NICER is not an imaging instrument, we cannot directly measure the background spectrum of each observation, and instead must use indirect methods to estimate or model it. We compare the effectiveness of the SCORPEON background model versus the best-performing background estimator 3C50 (as determined in \citealt{partington23}). A description of the SCORPEON background model is provided in Section~\ref{sec:spectralmodel}. The 3C50 background estimator divides each epoch into smaller Good Time Intervals (GTIs) of 10--120 s and constructs a background for each GTI based on instrumental parameters which represent out-of-focus photons and the presence of optical noise \citep{remillard_2022}. Each GTI is then filtered based on the net background-subtracted rate in the S0 (0.2--0.3 keV) and HGB (13--15 keV) bands, which are outside of NICER's source sensitivity band. We start our comparison with the same sample of 383 epochs of observations described in Section~\ref{sec:Data}. 

Given the faint nature of the soft excess spectral feature which is sensitive to background contamination, we filter the 3C50 spectra using the most stringent ``level 3" criterion: $|\mathrm{S0}_\mathrm{net}| < 0.2$ c s$^{-1}$ and $|\mathrm{HBG}_\mathrm{net}| < 0.05$ c s$^{-1}$. During 64 epochs, all GTIs are entirely screened out due to net count rates exceeding the filtering thresholds, resulting in only 319 epochs with successful background estimation by 3C50. We restrict our comparison to this set, and fit each spectrum to the source model of a power law plus a blackbody soft excess described in Section~\ref{sec:tiedsoftexcess}. The source and background count rates, exposure times, best-fit power law indices $\Gamma$, $\chi^2$, and degrees of freedom are listed in Table~\ref{table:AppendixASCORP} (SCORPEON) and Table~\ref{table:AppendixA3C50} (3C50). SCORPEON spectra are fit from 0.22--15 keV and 3C50 spectra are fit from 0.3--8 keV. 

\begin{figure}[t]
  \centering
  \includegraphics[width=\columnwidth]{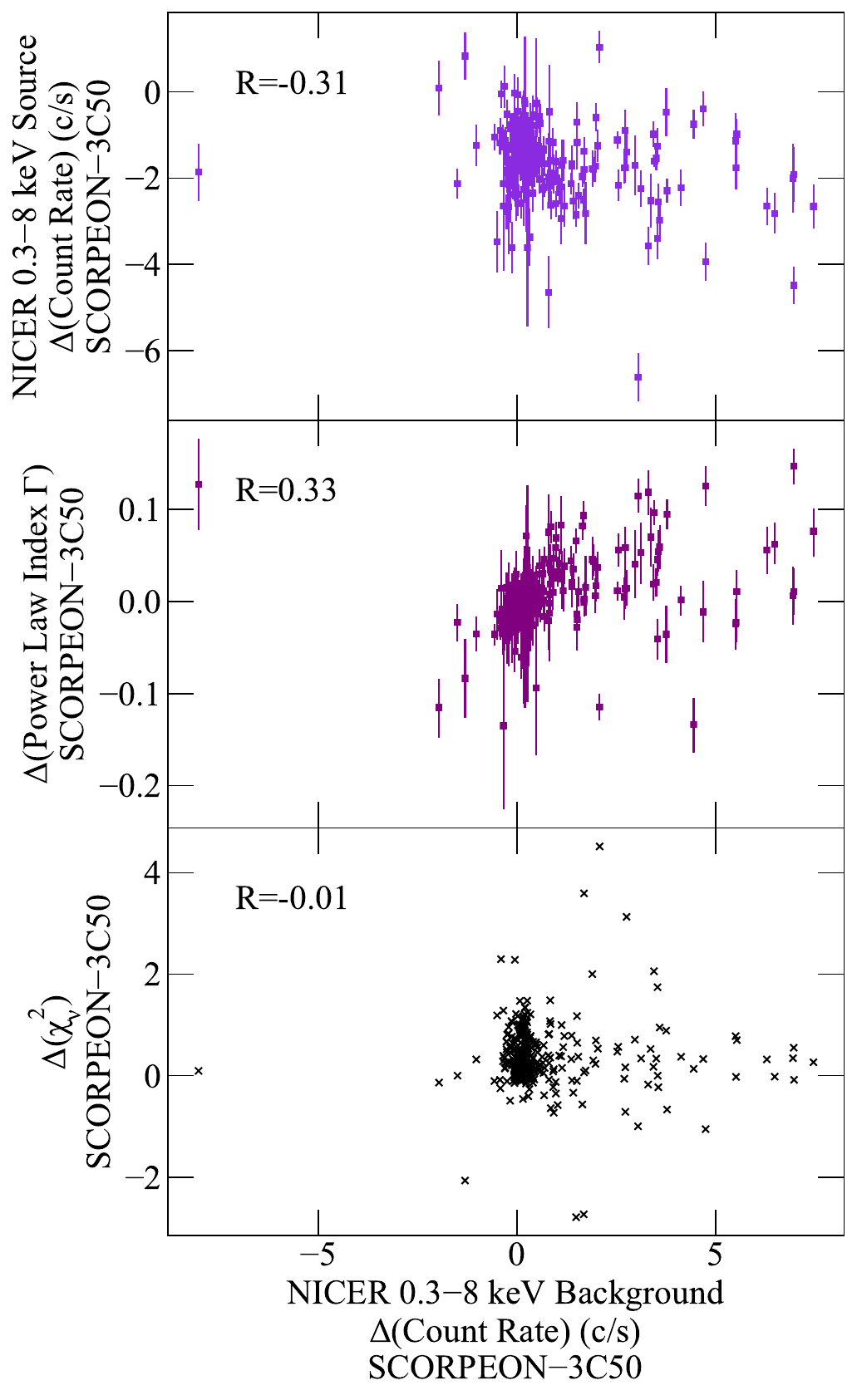}
  \caption{The difference ($\Delta$) between the 0.3--8 keV background rate measured by SCORPEON and 3C50, plotted versus the $\Delta$ in background-subtracted source count rate (Panel 1, light purple), power law spectral index $\Gamma$ (Panel 2, dark purple), and $\chi^2_{\nu}$ (Panel 3, black). The Pearson correlation coefficient $R$ is listed for each panel, demonstrating a lack of a significant relationship between the parameters.}
  \label{fig:AppendixAdeltabg}
\end{figure}

If either of the background estimators systematically underperform in a way which influences the shape or flux of the spectrum, we would expect to see a strong relationship between the difference in background count rates and the source spectrum parameters produced by the estimators. We demonstrate this in Figure~\ref{fig:AppendixAdeltabg} by comparing the differences in 0.3--8 keV background count rate to the difference in 0.3--8 keV source count rate (Pearson correlation $R=-0.31$), power law index $\Gamma$ ($R=0.33$), and $\chi^2_{\nu}$ ($R=-0.01$), finding no significant relationship. The distribution of the difference in background count rate is clustered about 0 with a tail of a few outliers reaching up to $\sim8$ c s$^{-1}$ (Fig.~\ref{fig:AppendixAdeltabg}), due to the GTI filtering process of 3C50. When the background rate is high, 3C50 is more likely to exclude the GTI, while SCORPEON includes the GTI and models the background. This process produces higher total background rates with SCORPEON without influencing the source parameters.

We compare the median 0.3--8 keV background-subtracted count rates of SCORPEON (31.2 c s$^{-1}$) and 3C50 (32.7 c s$^{-1}$) corresponding to a 5\% systematic uncertainty between the model and estimators. This is a substantial improvement over the 20\% systematic uncertainty between the 3C50, Space Weather, and Machine Learning estimators tested in \cite{partington23}. 

Given that the NICER background is characterized by a very flat spectrum, we would expect to see a sharp decrease in $\Gamma$ if the background was underestimated with a given method, contaminating the source spectra. We see no evidence of this, as the median difference in power law indices $\Delta(\Gamma)=0.00$ with a standard deviation across the sample of $\sigma=0.03$. The statistical uncertainty in the power law index reported in Table~\ref{table:uvlinkparam} is comparable to $\sigma$. 

Our analysis suggests that both SCORPEON and 3C50 provide a consistent characterization of the NICER background when it is well-behaved, e.g. in the absence of high-energy particle flares or intense optical loading. However, the total exposure time recovered across the 383 total epochs is 382.411 ks with SCORPEON, versus only 211.328 ks with 3C50. This shortage of 171.082 ks with 3C50 is caused by the rejection of GTIs outside of the Level 3 count rate thresholds, including the $\sim17$\% of epochs which were rejected in their entirety. Given the extreme improvement of SCORPEON over 3C50 in modeling the background during bright observing conditions (and thus recovering useful spectra for a greater number of epochs), we conduct our analysis using the SCORPEON background model.

\begin{deluxetable*}{ccccccc}
\tablewidth{0pt}
\tablecaption{\label{table:AppendixASCORP}SCORPEON Background Model: NICER background modeling and spectral analysis results with SCORPEON, using the source model from Section~\ref{sec:tiedsoftexcess}. All count rates are measured from 0.3--8 keV. The table covers a total of 319 epochs, corresponding to observations with a successful 3C50 background estimation. Reported uncertainties correspond to the 68\% ($1\sigma$) confidence interval for each parameter. The first five epochs are shown, with the full table available in machine-readable format.}
\tablehead{
\colhead{Obs. Date} &
\colhead{Source Rate} &
\colhead{Background Rate} &
\colhead{Exposure} &
\colhead{Power Law} &
\colhead{$\chi^2$} &
\colhead{Degrees of}\\
\colhead{(MJD)} &
\colhead{(counts~s$^{-1}$)} &
\colhead{(counts~s$^{-1}$)} &
\colhead{(s)} &
\colhead{Index $\Gamma$} &
\colhead{} &
\colhead{Freedom}
}
\startdata
58256.95 & $32.10\pm0.17$ & $1.10$ & $1652.00$ & $2.10\pm0.01$ & $115.28$ & $126$ \\ 
58257.00 & $31.14\pm0.22$ & $2.14$ & $809.00$ & $2.13\substack{+0.01\\ -0.02}$ & $111.59$ & $116$ \\ 
58272.38 & $37.25\pm0.20$ & $3.34$ & $1515.00$ & $2.16\pm0.01$ & $392.01$ & $158$ \\ 
58293.62 & $31.44\pm0.20$ & $4.24$ & $1103.00$ & $2.17\pm0.01$ & $166.26$ & $157$ \\ 
58294.65 & $31.56\pm0.18$ & $0.71$ & $1290.00$ & $2.13\pm0.01$ & $132.15$ & $98$ \\ 
... & ... & ... & ... & ... & ... & ... 
\enddata
\end{deluxetable*}

\begin{deluxetable*}{ccccccc}
\tablewidth{0pt}
\tablecaption{\label{table:AppendixA3C50}3C50 Background Estimator: NICER background estimation and source spectrum parameters using the 3C50 background. The source model is described in Section~\ref{sec:tiedsoftexcess}, and count rates are measured from 0.3--8 keV. The table includes 319 epochs. Reported uncertainties correspond to the 68\% ($1\sigma$) confidence interval for each parameter. The first five epochs are shown here, with the full table available in machine-readable format.}
\tablehead{
\colhead{Obs. Date} &
\colhead{Source Rate} &
\colhead{Background Rate} &
\colhead{Exposure} &
\colhead{Power Law} &
\colhead{$\chi^2$} &
\colhead{Degrees of}\\
\colhead{(MJD)} &
\colhead{(counts~s$^{-1}$)} &
\colhead{(counts~s$^{-1}$)} &
\colhead{(s)} &
\colhead{Index $\Gamma$} &
\colhead{} &
\colhead{Freedom}
}
\startdata
58256.96 & $33.70\pm0.18$ & $0.78$ & $1057.00$ & $2.08\pm0.01$ & $106.14$ & $141$ \\ 
58257.00 & $32.33\pm0.23$ & $1.73$ & $647.00$ & $2.12\pm0.01$ & $134.74$ & $136$ \\ 
58272.38 & $38.43\pm0.20$ & $1.83$ & $960.00$ & $2.17\pm0.01$ & $179.00$ & $137$ \\ 
58293.63 & $32.98\pm0.19$ & $0.73$ & $902.00$ & $2.15\pm0.01$ & $101.90$ & $141$ \\ 
58294.65 & $32.76\pm0.16$ & $0.58$ & $1266.00$ & $2.13\pm0.01$ & $171.03$ & $193$ \\ 
... & ... & ... & ... & ... & ... & ... 
\enddata
\end{deluxetable*}

\twocolumngrid
\section{Intercalibration of Spectral Shape with Swift and NICER}
\label{sec:swiftappendix}
\begin{figure*}[t]
  \centering
  \includegraphics[width=\textwidth]{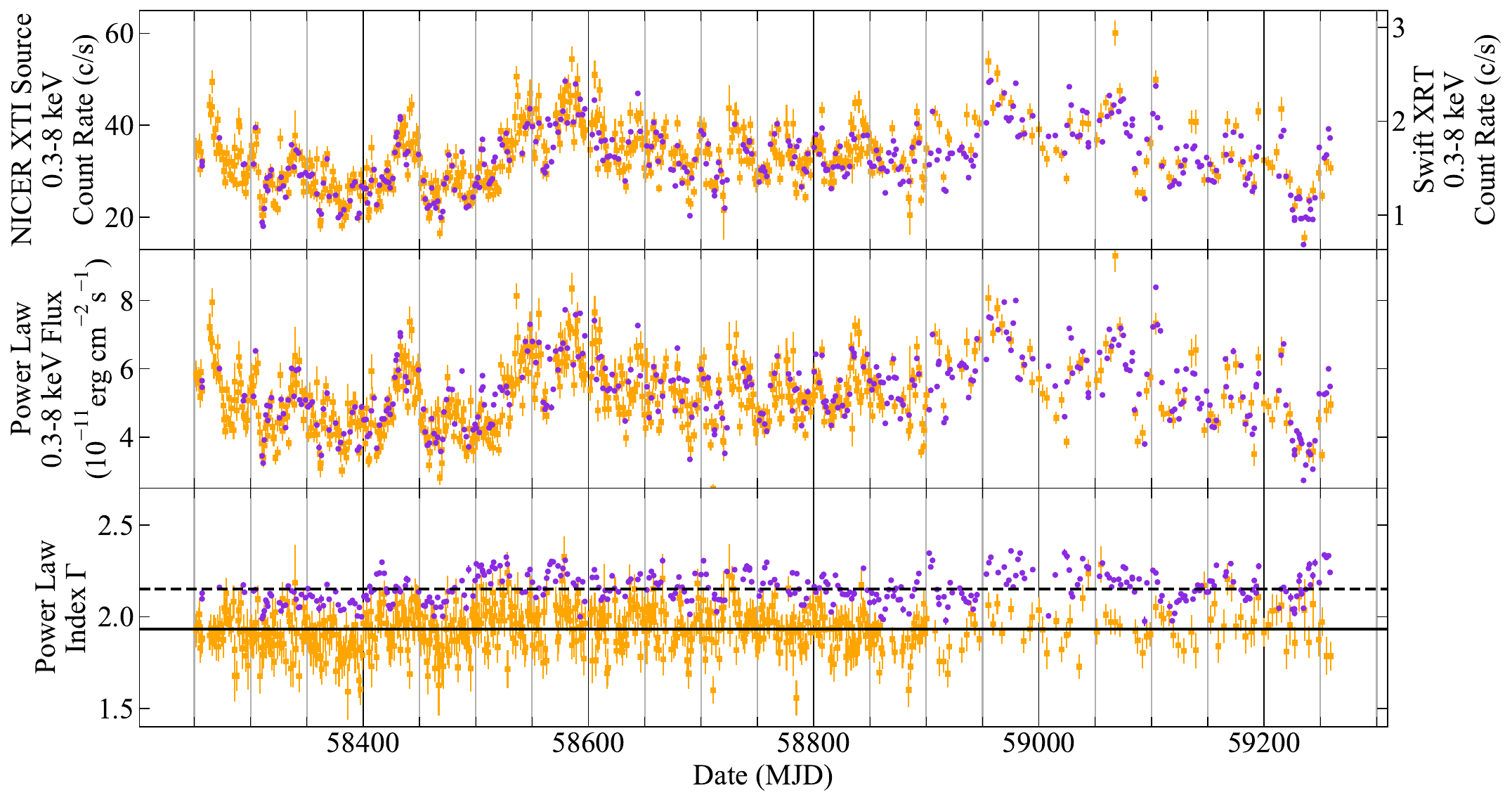}
  \caption{Light curves of count rate (Panel 1), power law flux (Panel 2) and power law index (Panel 3) measured for each epoch of observation with Swift XRT (orange squares) and NICER XTI (purple circles). The data reduction process is described in Appendix~\ref{sec:swiftappendix}. A significant offset in the measurement of $\Gamma$ by the two instruments is visible in Panel 3, indicated by the black lines which represent the sample median for NICER (dashed) and Swift (solid).}
  \label{fig:AppBrealspectra}
\end{figure*}

We observe a systematic offset in the shape of the power law spectrum (denoted by $\Gamma$) between observations with Swift XRT and NICER XTI, as shown in Panel 3 of Figure~\ref{fig:AppBrealspectra}. Swift spectra and light curves are generated using the XRT product builder (\citealt{evans07}, \citealt{evans09}). We fit spectra from each Swift epoch using the model described in Section~\ref{sec:tiedsoftexcess}, with the flux of the soft excess determined using Equation (1). The observed fluxes of the power law $\Phi_\mathrm{PL}$ are consistent between instruments. The Swift fluxes (Table~\ref{table:Apprealspectra}) have a median value of $\log(\Phi_\mathrm{PL,Swift})=-10.29$ (erg~cm$^{-2}$~s$^{-1}$) with a sample standard deviation of $\sigma=0.08$, and the median NICER flux (Table~\ref{table:uvlinkparam}) is $\log(\Phi_\mathrm{PL,NICER})=-10.27$ (erg~cm$^{-2}$~s$^{-1}$) with a standard deviation of $\sigma=0.08$. However, the median values for $\Gamma$ differ by more than $1\sigma$: $\Gamma_\mathrm{Swift}=1.9$ ($\sigma=0.1$) and $\Gamma_\mathrm{NICER}=2.15$ ($\sigma=0.08$). 

We test whether the difference in $\Gamma$ is caused by the Swift and NICER response matrices producing spectra with different slopes, or by a weighting bias when fitting due to NICER's increased count-rate sensitivity at soft energies compared to Swift. We simulate NICER and Swift spectra corresponding to each NICER epoch, using the model from Section~\ref{sec:tiedsoftexcess} with the best-fit parameters from Table~\ref{table:uvlinkparam} and the \textsc{fakeit} command in XSPEC. Each simulated NICER spectrum uses the real response, arf, and exposure time from its associated epoch. 

To compare Swift and NICER count rates, we simulated Swift spectra corresponding to each NICER epoch using the response and arf from the first Swift observation (OBSID ``00094060001"). A consistent response matrix is required to provide accurate count rate variability, since the XRT Spectrum Builder \citep{evans09} weights the combined spectrum of all snapshots in each epoch by counts, rather than exposure time. Since the snapshot lengths are variable, the combined spectrum will have the correct spectral shape, but the total count rate will be different than the one determined by the XRT Light Curve Builder \citep{evans07}.

\begin{figure*}[t]
  \centering
  \includegraphics[width=\textwidth]{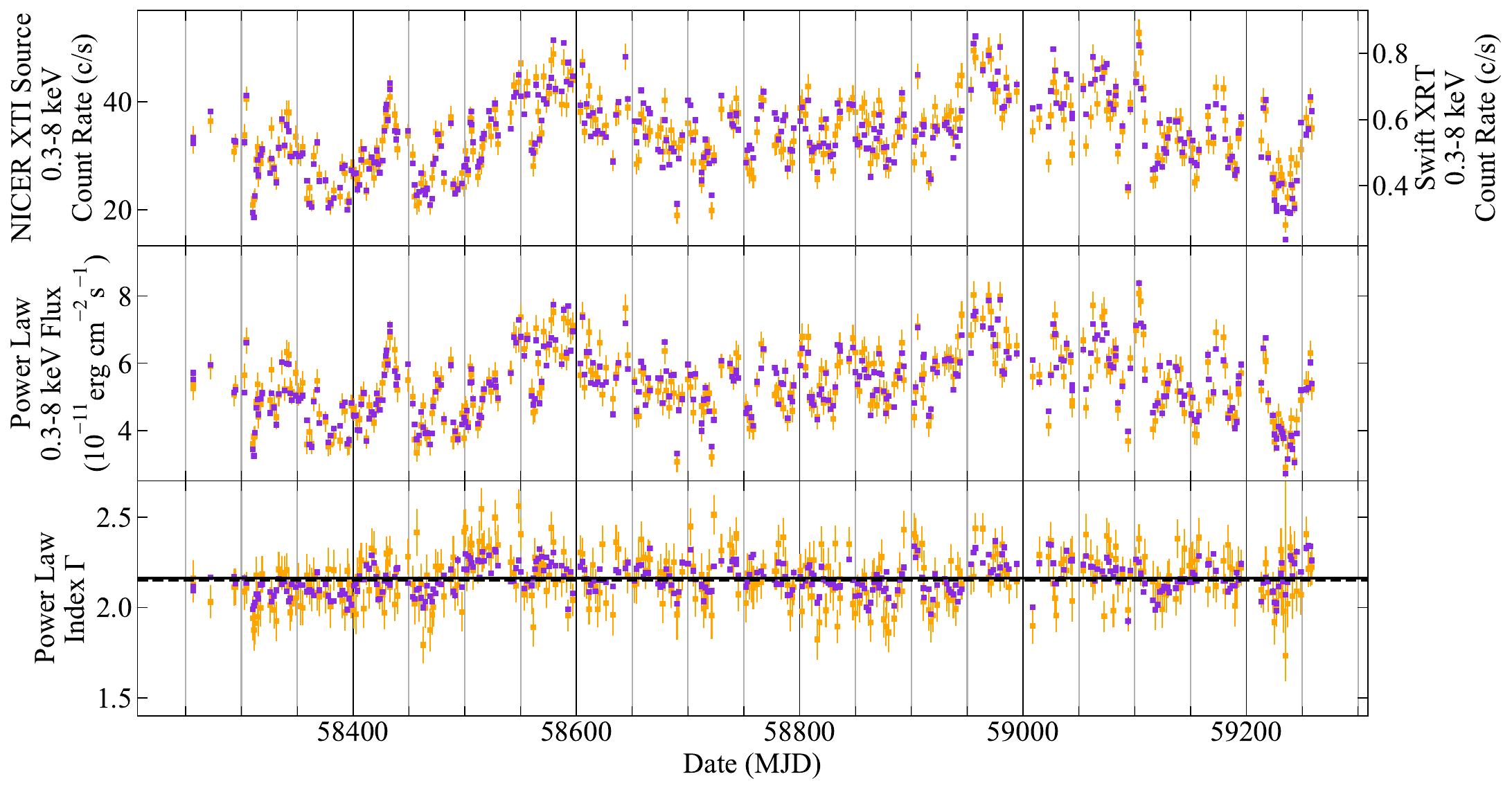}
  \caption{Same as Figure~\ref{fig:AppBrealspectra}, but for simulated spectra using the \textsc{fakeit} command in XSPEC corresponding to the best-fit model from each NICER epoch, as described in Appendix~\ref{sec:swiftappendix}. Swift XRT data are shown as orange squares, and NICER XTI data are shown as purple circles. The black lines in Panel 3 show the median values of $\Gamma$ for NICER (dashed) and Swift (solid). Despite the differing response matrices for each instrument, the measured $\Gamma$ values agree, in contrast to the real data in Figure~\ref{fig:AppBrealspectra}.}
  \label{fig:AppBfakespectra}
\end{figure*}

We fit each simulated Swift and NICER spectrum to the same model as the real spectra, with NICER results reported in Table~\ref{table:AppNICERfakespectra} and Swift results in Table~\ref{table:Appswiftfakespectra}. The comparison of $\Gamma$ in Panel 3 of Figure~\ref{fig:AppBfakespectra} clearly shows that the difference in sensitivity to soft X-rays does not cause the offset in $\Gamma$ seen in the real spectra (Fig.~\ref{fig:AppBrealspectra}). The median fluxes agree, as do the power law indices: $\log(\Phi_\mathrm{PL,Swift,simulated})=-10.28$ with a sample standard deviation of $\sigma=0.08$, $\log(\Phi_\mathrm{PL,NICER,simulated})=-10.27$ ($\sigma=0.08$), $\Gamma_\mathrm{Swift,simulated}=2.2$ ($\sigma=0.1$), and  $\Gamma_\mathrm{NICER,simulated}=2.15$ ($\sigma=0.08$). This indicates that the observed offset in $\Gamma$ between Swift and NICER is intrinsic to a difference in calibration between the instruments. This should be taken into consideration when comparing \textit{Swift} and NICER for AGN-like spectra, and addressed in future updates to the NICER response matrix. 

\begin{deluxetable*}{ccccccc}
\tablewidth{0pt}
\tablecaption{\label{table:Apprealspectra}Swift Spectral Analysis: Count rate and best-fit spectral parameters for each real Swift XRT epoch. Reported uncertainties correspond to the 68\% ($1\sigma$) confidence interval for each parameter. Count rates are from the Swift XRT Light Curve Builder \citep{evans07} and spectra are generated using the Spectrum Builder \citep{evans09}. The first five rows are presented, with the rest available in machine-readable format.}
\tablehead{
\colhead{Obs. Date} &
\colhead{Count Rate} &
\colhead{Power Law} &
\colhead{Power Law} &
\colhead{Soft Excess} &
\colhead{Best fit} &
\colhead{Degrees}\\
\colhead{(MJD)} &
\colhead{(counts~s$^{-1}$)} &
\colhead{Flux $\log(\Phi_{\mathrm{PL}})$} &
\colhead{Index $\Gamma$} &
\colhead{Flux $\log(\Phi_{\mathrm{SE}})$} &
\colhead{C-Statistic} &
\colhead{of}\\
\colhead{} &
\colhead{0.3-8 keV} &
\colhead{(erg~cm$^{-2}$~s$^{-1}$)} &
\colhead{} &
\colhead{(erg~cm$^{-2}$~s$^{-1}$)} &
\colhead{} &
\colhead{Freedom}\\
}
\startdata
58251.66 & $1.71\pm0.11$ & $-10.23\pm0.03$ & $1.9\pm0.1$ & $-11.781\pm0.008$ & $40.36$ & $39$ \\ 
58252.46 & $1.69\pm0.10$ & $-10.24\pm0.03$ & $1.9\pm0.1$ & $-11.775\pm0.008$ & $32.81$ & $92$ \\ 
58253.86 & $1.71\pm0.10$ & $-10.24\pm0.03$ & $2.0\pm0.1$ & $-11.749\pm0.008$ & $37.71$ & $42$ \\ 
58254.58 & $1.76\pm0.10$ & $-10.25\pm0.03$ & $2.0\pm0.1$ & $-11.759\pm0.008$ & $46.36$ & $39$ \\ 
58255.52 & $1.49\pm0.08$ & $-10.29\pm0.03$ & $1.8\pm0.1$ & $-11.751\pm0.008$ & $62.90$ & $41$ \\ 
... & ... & ... & ... & ... & ... & ...
\enddata
\end{deluxetable*}

\begin{deluxetable*}{ccccccc}
\tablewidth{0pt}
\tablecaption{\label{table:AppNICERfakespectra}NICER Simulated Spectra: Count rate and spectral parameters for each simulated NICER XTI epoch. The production of simulated spectra is described in Appendix~\ref{sec:swiftappendix}. Reported uncertainties correspond to the 68\% ($1\sigma$) confidence interval for each parameter. The first five rows are presented, with the rest available in machine-readable format.}
\tablehead{
\colhead{Obs. Date} &
\colhead{Count Rate} &
\colhead{Power Law} &
\colhead{Power Law} &
\colhead{Soft Excess} &
\colhead{Best fit} &
\colhead{Degrees}\\
\colhead{(MJD)} &
\colhead{(counts~s$^{-1}$)} &
\colhead{Flux $\log(\Phi_{\mathrm{PL}})$} &
\colhead{Index $\Gamma$} &
\colhead{Flux $\log(\Phi_{\mathrm{SE}})$} &
\colhead{$\chi^2_{\nu}$} &
\colhead{of}\\
\colhead{} &
\colhead{0.3-8 keV} &
\colhead{(erg~cm$^{-2}$~s$^{-1}$)} &
\colhead{} &
\colhead{(erg~cm$^{-2}$~s$^{-1}$)} &
\colhead{} &
\colhead{Freedom}\\
}
\startdata
58256.95 & $33.40\pm0.17$ & $-10.243\pm0.003$ & $2.09\pm0.01$ & $-11.739\pm0.008$ & $94.85$ & $98$ \\ 
58257.00 & $32.31\pm0.23$ & $-10.258\pm0.004$ & $2.11\pm0.01$ & $-11.739\pm0.008$ & $102.87$ & $95$ \\ 
58272.38 & $38.23\pm0.20$ & $-10.225\pm0.003$ & $2.17\pm0.01$ & $-11.619\pm0.008$ & $91.36$ & $102$ \\ 
58293.62 & $32.83\pm0.21$ & $-10.290\substack{+0.003\\ -0.004}$ & $2.17\pm0.01$ & $-11.738\pm0.008$ & $79.71$ & $102$ \\ 
58294.65 & $32.66\pm0.19$ & $-10.283\pm0.003$ & $2.13\pm0.01$ & $-11.760\pm0.008$ & $119.48$ & $98$ \\ 
... & ... & ... & ... & ... & ... & ...
\enddata
\end{deluxetable*}

\begin{deluxetable*}{ccccccc}
\tablewidth{0pt}
\tablecaption{\label{table:Appswiftfakespectra}Swift Simulated Spectra: Count rate and spectral parameters for each simulated Swift XRT observation based on the best-fit spectral parameters from each real NICER XTI epoch. The production of simulated spectra is described in Appendix~\ref{sec:swiftappendix}. Reported uncertainties correspond to the 68\% ($1\sigma$) confidence interval for each parameter. The first five rows are presented, with the rest available in machine-readable format.}
\tablehead{
\colhead{Obs. Date} &
\colhead{Count Rate} &
\colhead{Power Law} &
\colhead{Power Law} &
\colhead{Soft Excess} &
\colhead{Best fit} &
\colhead{Degrees}\\
\colhead{(MJD)} &
\colhead{(counts~s$^{-1}$)} &
\colhead{Flux $\log(\Phi_{\mathrm{PL}})$} &
\colhead{Index $\Gamma$} &
\colhead{Flux $\log(\Phi_{\mathrm{SE}})$} &
\colhead{C-Statistic} &
\colhead{of}\\
\colhead{} &
\colhead{0.3-8 keV} &
\colhead{(erg~cm$^{-2}$~s$^{-1}$)} &
\colhead{} &
\colhead{(erg~cm$^{-2}$~s$^{-1}$)} &
\colhead{} &
\colhead{Freedom}\\
}
\startdata
58256.95 & $0.55\pm0.03$ & $-10.27\pm0.03$ & $2.1\pm0.1$ & $-11.739\pm0.008$ & $31.66$ & $37$ \\ 
58257.00 & $0.53\pm0.03$ & $-10.28\pm0.03$ & $2.2\pm0.1$ & $-11.739\pm0.008$ & $22.60$ & $150$ \\ 
58272.38 & $0.59\pm0.03$ & $-10.23\pm0.03$ & $2.0\pm0.1$ & $-11.619\pm0.008$ & $30.23$ & $210$ \\ 
58293.62 & $0.50\pm0.03$ & $-10.30\pm0.03$ & $2.1\pm0.1$ & $-11.738\pm0.008$ & $29.98$ & $94$ \\ 
58294.65 & $0.52\pm0.03$ & $-10.28\pm0.03$ & $2.1\pm0.1$ & $-11.760\pm0.008$ & $42.25$ & $242$ \\ 
... & ... & ... & ... & ... & ... & ...
\enddata
\end{deluxetable*}

\newpage
\bibliographystyle{apj}
\bibliography{ref.bib}

\begin{thebibliography}{}
\expandafter\ifx\csname natexlab\endcsname\relax\def\natexlab#1{#1}\fi

\bibitem[{{Arnaud}(1996)}]{arnaud96}
{Arnaud}, K.~A. 1996, in Astronomical Society of the Pacific Conference Series, Vol. 101, Astronomical Data Analysis Software and Systems V, ed. G.~H. {Jacoby} \& J.~{Barnes}, 17

\bibitem[{Arévalo {et~al.}(2005)Arévalo, Papadakis, Kuhlbrodt, \& Brinkmann}]{Arevalo_2005}
Arévalo, P., Papadakis, I., Kuhlbrodt, B., \& Brinkmann, W. 2005, A\&A, 430, 435

\bibitem[{Arévalo \& Uttley(2006)}]{Arevalo_2006}
Arévalo, P., \& Uttley, P. 2006, \mnras, 367, 801

\bibitem[{Arévalo {et~al.}(2008)Arévalo, Uttley, Kaspi, Breedt, Lira, \& McHardy}]{Arevalo_2008}
Arévalo, P., Uttley, P., Kaspi, S., {et~al.} 2008, \mnras, 389, 1479

\bibitem[{{Ballantyne} \& {Xiang}(2020)}]{Ballantyne_2020}
{Ballantyne}, D.~R., \& {Xiang}, X. 2020, \mnras, 496, 4255

\bibitem[{{Blackburn}(1995)}]{Blackburn_1995}
{Blackburn}, J.~K. 1995, in Astronomical Society of the Pacific Conference Series, Vol.~77, Astronomical Data Analysis Software and Systems IV, ed. R.~A. {Shaw}, H.~E. {Payne}, \& J.~J.~E. {Hayes}, 367

\bibitem[{{Boissay} {et~al.}(2016){Boissay}, {Ricci}, \& {Paltani}}]{boissay16}
{Boissay}, R., {Ricci}, C., \& {Paltani}, S. 2016, \aap, 588, A70

\bibitem[{{Cackett} {et~al.}(2021){Cackett}, {Bentz}, \& {Kara}}]{cackett21}
{Cackett}, E.~M., {Bentz}, M.~C., \& {Kara}, E. 2021, iScience, 24, 102557

\bibitem[{{Cackett} {et~al.}(2007){Cackett}, {Horne}, \& {Winkler}}]{Cackett2007}
{Cackett}, E.~M., {Horne}, K., \& {Winkler}, H. 2007, \mnras, 380, 669

\bibitem[{Cackett {et~al.}(2014)Cackett, Zoghbi, Reynolds, Fabian, Kara, Uttley, \& Wilkins}]{cackett14}
Cackett, E.~M., Zoghbi, A., Reynolds, C., {et~al.} 2014, \mnras, 438, 2980

\bibitem[{Cackett {et~al.}(2023)Cackett, Gelbord, Barth, Rosa, Edelson, Goad, Homayouni, Horne, Kara, Kriss, Korista, Landt, Plesha, Arav, Bentz, Boizelle, Bontà, Dehghanian, Donnan, Du, Ferland, Fian, Filippenko, Buitrago, Grier, Hall, Hu, Ilić, Kaastra, Kaspi, Kochanek, Kovačević, Kynoch, Li, McLane, Mehdipour, Miller, Montano, Netzer, Panagiotou, Partington, Popović, Proga, Rogantini, Sanmartim, Siebert, Storchi-Bergmann, Vestergaard, Wang, Waters, \& Zaidouni}]{Cackett_2023}
Cackett, E.~M., Gelbord, J., Barth, A.~J., {et~al.} 2023, \apj, 958, 195

\bibitem[{{Cardelli} {et~al.}(1989){Cardelli}, {Clayton}, \& {Mathis}}]{Cardelli_1989}
{Cardelli}, J.~A., {Clayton}, G.~C., \& {Mathis}, J.~S. 1989, \apj, 345, 245

\bibitem[{Crummy {et~al.}(2006)Crummy, Fabian, Gallo, \& Ross}]{Crummy_2006}
Crummy, J., Fabian, A.~C., Gallo, L., \& Ross, R.~R. 2006, MNRAS, 365, 1067

\bibitem[{{De Marco} {et~al.}(2013){De Marco}, {Ponti}, {Cappi}, {Dadina}, {Uttley}, {Cackett}, {Fabian}, \& {Miniutti}}]{demarco13}
{De Marco}, B., {Ponti}, G., {Cappi}, M., {et~al.} 2013, \mnras, 431, 2441

\bibitem[{{Done} {et~al.}(2012){Done}, {Davis}, {Jin}, {Blaes}, \& {Ward}}]{Done12}
{Done}, C., {Davis}, S.~W., {Jin}, C., {Blaes}, O., \& {Ward}, M. 2012, \mnras, 420, 1848

\bibitem[{{Donnan} {et~al.}(2021){Donnan}, {Horne}, \& {Hern{\'a}ndez Santisteban}}]{Donnan_2021}
{Donnan}, F.~R., {Horne}, K., \& {Hern{\'a}ndez Santisteban}, J.~V. 2021, \mnras, 508, 5449

\bibitem[{Edelson {et~al.}(2017)Edelson, Gelbord, Cackett, Connolly, Done, Fausnaugh, Gardner, Gehrels, Goad, Horne, McHardy, Peterson, Vaughan, Vestergaard, Breeveld, Barth, Bentz, Bottorff, Brandt, Crawford, Bont{\`{a}}, Emmanoulopoulos, Evans, Jaimes, Filippenko, Ferland, Grupe, Joner, Kennea, Korista, Krimm, Kriss, Leonard, Mathur, Netzer, Nousek, Page, Romero-Colmenero, Siegel, Starkey, Treu, Vogler, Winkler, \& Zheng}]{edelson2017}
Edelson, R., Gelbord, J., Cackett, E., {et~al.} 2017, ApJ, 840, 41

\bibitem[{{Edelson} {et~al.}(2019){Edelson}, {Gelbord}, {Cackett}, {Peterson}, {Horne}, {Barth}, {Starkey}, {Bentz}, {Brandt}, {Goad}, {Joner}, {Korista}, {Netzer}, {Page}, {Uttley}, {Vaughan}, {Breeveld}, {Cenko}, {Done}, {Evans}, {Fausnaugh}, {Ferland}, {Gonzalez-Buitrago}, {Gropp}, {Grupe}, {Kaastra}, {Kennea}, {Kriss}, {Mathur}, {Mehdipour}, {Mudd}, {Nousek}, {Schmidt}, {Vestergaard}, \& {Villforth}}]{edelson19}
{Edelson}, R., {Gelbord}, J., {Cackett}, E., {et~al.} 2019, \apj, 870, 123

\bibitem[{Emmanoulopoulos {et~al.}(2014)Emmanoulopoulos, Papadakis, Dovčiak, \& McHardy}]{Emmanoulopoulos14}
Emmanoulopoulos, D., Papadakis, I.~E., Dovčiak, M., \& McHardy, I.~M. 2014, \mnras, 439, 3931

\bibitem[{Emmanoulopoulos {et~al.}(2011)Emmanoulopoulos, Papadakis, McHardy, Nicastro, Bianchi, \& Arévalo}]{Emmanoulopoulos11}
Emmanoulopoulos, D., Papadakis, I.~E., McHardy, I.~M., {et~al.} 2011, \mnras, 415, 1895

\bibitem[{{Evans} {et~al.}(2007){Evans}, {Beardmore}, {Page}, {Tyler}, {Osborne}, {Goad}, {O'Brien}, {Vetere}, {Racusin}, {Morris}, {Burrows}, {Capalbi}, {Perri}, {Gehrels}, \& {Romano}}]{evans07}
{Evans}, P.~A., {Beardmore}, A.~P., {Page}, K.~L., {et~al.} 2007, \aap, 469, 379

\bibitem[{{Evans} {et~al.}(2009){Evans}, {Beardmore}, {Page}, {Osborne}, {O'Brien}, {Willingale}, {Starling}, {Burrows}, {Godet}, {Vetere}, {Racusin}, {Goad}, {Wiersema}, {Angelini}, {Capalbi}, {Chincarini}, {Gehrels}, {Kennea}, {Margutti}, {Morris}, {Mountford}, {Pagani}, {Perri}, {Romano}, \& {Tanvir}}]{evans09}
---. 2009, \mnras, 397, 1177

\bibitem[{{Fabian} {et~al.}(2015){Fabian}, {Lohfink}, {Kara}, {Parker}, {Vasudevan}, \& {Reynolds}}]{fabian15}
{Fabian}, A.~C., {Lohfink}, A., {Kara}, E., {et~al.} 2015, \mnras, 451, 4375

\bibitem[{Fabian {et~al.}(2014)Fabian, Parker, Wilkins, Miller, Kara, Reynolds, \& Dauser}]{fabian14}
Fabian, A.~C., Parker, M.~L., Wilkins, D.~R., {et~al.} 2014, \mnras, 439, 2307

\bibitem[{{Fausnaugh} {et~al.}(2016){Fausnaugh}, {Denney}, {Barth}, {Bentz}, {Bottorff}, {Carini}, {Croxall}, {De Rosa}, {Goad}, {Horne}, {Joner}, {Kaspi}, {Kim}, {Klimanov}, {Kochanek}, {Leonard}, {Netzer}, {Peterson}, {Schn{\"u}lle}, {Sergeev}, {Vestergaard}, {Zheng}, {Zu}, {Anderson}, {Ar{\'e}valo}, {Bazhaw}, {Borman}, {Boroson}, {Brandt}, {Breeveld}, {Brewer}, {Cackett}, {Crenshaw}, {Dalla Bont{\`a}}, {De Lorenzo-C{\'a}ceres}, {Dietrich}, {Edelson}, {Efimova}, {Ely}, {Evans}, {Filippenko}, {Flatland}, {Gehrels}, {Geier}, {Gelbord}, {Gonzalez}, {Gorjian}, {Grier}, {Grupe}, {Hall}, {Hicks}, {Horenstein}, {Hutchison}, {Im}, {Jensen}, {Jones}, {Kaastra}, {Kelly}, {Kennea}, {Kim}, {Korista}, {Kriss}, {Lee}, {Lira}, {MacInnis}, {Manne-Nicholas}, {Mathur}, {McHardy}, {Montouri}, {Musso}, {Nazarov}, {Norris}, {Nousek}, {Okhmat}, {Pancoast}, {Papadakis}, {Parks}, {Pei}, {Pogge}, {Pott}, {Rafter}, {Rix}, {Saylor}, {Schimoia}, {Siegel}, {Spencer}, {Starkey}, {Sung}, {Teems}, {Treu}, {Turner}, {Uttley}, {Villforth},
  {Weiss}, {Woo}, {Yan}, \& {Young}}]{fausnaugh16}
{Fausnaugh}, M.~M., {Denney}, K.~D., {Barth}, A.~J., {et~al.} 2016, \apj, 821, 56

\bibitem[{{Frank} {et~al.}(2002){Frank}, {King}, \& {Raine}}]{Frank02}
{Frank}, J., {King}, A., \& {Raine}, D.~J. 2002, {Accretion Power in Astrophysics: Third Edition}

\bibitem[{{Gierli{\'n}ski} \& {Done}(2004)}]{gierlinski04}
{Gierli{\'n}ski}, M., \& {Done}, C. 2004, \mnras, 349, L7

\bibitem[{{Haardt} \& {Maraschi}(1991)}]{Haardt_1991}
{Haardt}, F., \& {Maraschi}, L. 1991, \apjl, 380, L51

\bibitem[{Hagen \& Done(2023)}]{hagendone23}
Hagen, S., \& Done, C. 2023, \mnras, 521, 251

\bibitem[{Hancock {et~al.}(2022)Hancock, Young, \& Chainakun}]{Hancock22}
Hancock, S., Young, A.~J., \& Chainakun, P. 2022, \mnras, 514, 5403

\bibitem[{{Hern{\'a}ndez Santisteban} {et~al.}(2020){Hern{\'a}ndez Santisteban}, {Edelson}, {Horne}, {Gelbord}, {Barth}, {Cackett}, {Goad}, {Netzer}, {Starkey}, {Uttley}, {Brandt}, {Korista}, {Lohfink}, {Onken}, {Page}, {Siegel}, {Vestergaard}, {Bisogni}, {Breeveld}, {Cenko}, {Dalla Bont{\`a}}, {Evans}, {Ferland}, {Gonzalez-Buitrago}, {Grupe}, {Joner}, {Kriss}, {LaPorte}, {Mathur}, {Marshall}, {Mehdipour}, {Mudd}, {Peterson}, {Schmidt}, {Vaughan}, \& {Valenti}}]{Hernandez2020}
{Hern{\'a}ndez Santisteban}, J.~V., {Edelson}, R., {Horne}, K., {et~al.} 2020, \mnras, 498, 5399

\bibitem[{Hunter(2007)}]{Hunter_2007}
Hunter, J.~D. 2007, Computing in Science \& Engineering, 9, 90

\bibitem[{{Kaastra} \& {Bleeker}(2016)}]{kaastra16}
{Kaastra}, J.~S., \& {Bleeker}, J.~A.~M. 2016, \aap, 587, A151

\bibitem[{{Kara} {et~al.}(2016){Kara}, {Alston}, {Fabian}, {Cackett}, {Uttley}, {Reynolds}, \& {Zoghbi}}]{kara16}
{Kara}, E., {Alston}, W.~N., {Fabian}, A.~C., {et~al.} 2016, \mnras, 462, 511

\bibitem[{Kara {et~al.}(2023)Kara, Barth, Cackett, Gelbord, Montano, Li, Santana, Horne, Alston, Buisson, Chelouche, Du, Fabian, Fian, Gallo, Goad, Grupe, Buitrago, Santisteban, Kaspi, Hu, Komossa, Kriss, Lewin, Lewis, Loewenstein, Lohfink, Masterson, McHardy, Mehdipour, Miller, Panagiotou, Parker, Pinto, Remillard, Reynolds, Rogantini, Wang, Wang, \& Wilkins}]{Kara_2023}
Kara, E., Barth, A.~J., Cackett, E.~M., {et~al.} 2023, \apj, 947, 62

\bibitem[{Kubota \& Done(2018)}]{kubota18}
Kubota, A., \& Done, C. 2018, \mnras, 480, 1247

\bibitem[{Laor {et~al.}(1997)Laor, Fiore, Elvis, Wilkes, \& McDowell}]{Laor_1997}
Laor, A., Fiore, F., Elvis, M., Wilkes, B.~J., \& McDowell, J.~C. 1997, \apj, 477, 93

\bibitem[{Lohfink {et~al.}(2016)Lohfink, Reynolds, Pinto, Alston, Boggs, Christensen, Craig, Fabian, Hailey, Harrison, Kara, Matt, Parker, Stern, Walton, \& Zhang}]{Lohfink_2016}
Lohfink, A.~M., Reynolds, C.~S., Pinto, C., {et~al.} 2016, \apj, 821, 11

\bibitem[{{Lubi{\'n}ski} {et~al.}(2016){Lubi{\'n}ski}, {Beckmann}, {Gibaud}, {Paltani}, {Papadakis}, {Ricci}, {Soldi}, {T{\"u}rler}, {Walter}, \& {Zdziarski}}]{lubinski16}
{Lubi{\'n}ski}, P., {Beckmann}, V., {Gibaud}, L., {et~al.} 2016, \mnras, 458, 2454

\bibitem[{Magdziarz {et~al.}(1998)Magdziarz, Blaes, Zdziarski, Johnson, \& Smith}]{Magdziarz_1998}
Magdziarz, P., Blaes, O.~M., Zdziarski, A.~A., Johnson, W.~N., \& Smith, D.~A. 1998, \mnras, 301, 179

\bibitem[{{Mahmoud} \& {Done}(2020)}]{mahmoud20}
{Mahmoud}, R.~D., \& {Done}, C. 2020, \mnras, 491, 5126

\bibitem[{Mahmoud {et~al.}(2022)Mahmoud, Done, Porquet, \& Lobban}]{mahmoud22}
Mahmoud, R.~D., Done, C., Porquet, D., \& Lobban, A. 2022, \mnras, 521, 3585

\bibitem[{{Matt} {et~al.}(2014){Matt}, {Marinucci}, {Guainazzi}, {Brenneman}, {Elvis}, {Lohfink}, {Ar{\`e}valo}, {Boggs}, {Cappi}, {Christensen}, {Craig}, {Fabian}, {Fuerst}, {Hailey}, {Harrison}, {Parker}, {Reynolds}, {Stern}, {Walton}, \& {Zhang}}]{matt14}
{Matt}, G., {Marinucci}, A., {Guainazzi}, M., {et~al.} 2014, \mnras, 439, 3016

\bibitem[{Mehdipour {et~al.}(2023)Mehdipour, Kriss, Kaastra, Costantini, \& Mao}]{Mehdipour_2023}
Mehdipour, M., Kriss, G.~A., Kaastra, J.~S., Costantini, E., \& Mao, J. 2023, \apjl, 952, L5

\bibitem[{{Mehdipour} {et~al.}(2011){Mehdipour}, {Branduardi-Raymont}, {Kaastra}, {Petrucci}, {Kriss}, {Ponti}, {Blustin}, {Paltani}, {Cappi}, {Detmers}, \& {Steenbrugge}}]{mehdipour11}
{Mehdipour}, M., {Branduardi-Raymont}, G., {Kaastra}, J.~S., {et~al.} 2011, \aap, 534, A39

\bibitem[{{Mehdipour} {et~al.}(2015){Mehdipour}, {Kaastra}, {Kriss}, {Cappi}, {Petrucci}, {Steenbrugge}, {Arav}, {Behar}, {Bianchi}, {Boissay}, {Branduardi-Raymont}, {Costantini}, {Ebrero}, {Di Gesu}, {Harrison}, {Kaspi}, {De Marco}, {Matt}, {Paltani}, {Peterson}, {Ponti}, {Pozo Nu{\~n}ez}, {De Rosa}, {Ursini}, {de Vries}, {Walton}, \& {Whewell}}]{Mehdipour2015}
{Mehdipour}, M., {Kaastra}, J.~S., {Kriss}, G.~A., {et~al.} 2015, \aap, 575, A22

\bibitem[{{Middei} {et~al.}(2023){Middei}, {Petrucci, P.-O.}, {Bianchi, S.}, {Ursini, F.}, {Matzeu, G. A.}, {Vagnetti, F.}, {Tortosa, A.}, {Marinucci, A.}, {Matt, G.}, {Piconcelli, E.}, {De Rosa, A.}, {De Marco, B.}, {Reeves, J.}, {Perri, M.}, {Guainazzi, M.}, {Cappi, M.}, \& {Done, C.}}]{middei23}
{Middei}, R., {Petrucci, P.-O.}, {Bianchi, S.}, {et~al.} 2023, A\&A, 672, A101

\bibitem[{{Neustadt} \& {Kochanek}(2022)}]{Neustadt_2022}
{Neustadt}, J.~M.~M., \& {Kochanek}, C.~S. 2022, \mnras, 513, 1046

\bibitem[{Panagiotou {et~al.}(2022{\natexlab{a}})Panagiotou, Kara, \& Dovčiak}]{Panagiotou_2022_xraycorrel}
Panagiotou, C., Kara, E., \& Dovčiak, M. 2022{\natexlab{a}}, \apj, 941, 57

\bibitem[{Panagiotou {et~al.}(2022{\natexlab{b}})Panagiotou, Papadakis, Kara, Kammoun, \& Dovčiak}]{Panagiotou_2022_5548}
Panagiotou, C., Papadakis, I., Kara, E., Kammoun, E., \& Dovčiak, M. 2022{\natexlab{b}}, \apj, 935, 93

\bibitem[{Partington {et~al.}(2023)Partington, Cackett, Kara, Kriss, Barth, Rosa, Homayouni, Horne, Landt, Zoghbi, Edelson, Arav, Boizelle, Bentz, Brotherton, Byun, Bontà, Dehghanian, Du, Fian, Filippenko, Gelbord, Goad, Buitrago, Grier, Hall, Hu, Ilić, Joner, Kaspi, Kochanek, Korista, Kovačević, Kynoch, McLane, Mehdipour, Miller, Panagiotou, Plesha, Popović, Proga, Rogantini, Storchi-Bergmann, Sanmartim, Siebert, Vestergaard, Ward, Waters, \& Zaidouni}]{partington23}
Partington, E.~R., Cackett, E.~M., Kara, E., {et~al.} 2023, \apj, 947, 2

\bibitem[{Pascual {et~al.}(1997)Pascual, Alloin, Clavel, Crenshaw, Horne, Kriss, Krolik, Malkan, Netzer, O'Brien, Peterson, Reichert, Wamsteker, Alexander, Barr, Blandford, Bregman, Carone, Clements, Courvoisier, Robertis, Dietrich, Dottori, Edelson, Filippenko, Gaskell, Huchra, Hutchings, Kollatschny, Koratkar, Korista, Laor, MacAlpine, Martin, Maoz, McCollum, Morris, Perola, Pogge, Ptak, González, Espinoza, Rokaki, Lleó, Sekiguchi, Shull, Snijders, Sparke, Stirpe, Stoner, Sun, Wagner, Wanders, Wilkes, Winge, \& Zheng}]{Rodriguez_1997}
Pascual, P. M.~R., Alloin, D., Clavel, J., {et~al.} 1997, ApJ Supplement Series, 110, 9

\bibitem[{Peterson {et~al.}(1998)Peterson, Wanders, Horne, Collier, Alexander, Kaspi, \& Maoz}]{Peterson_1998}
Peterson, B.~M., Wanders, I., Horne, K., {et~al.} 1998, \pasp, 110, 660

\bibitem[{{Peterson} {et~al.}(2004){Peterson}, {Ferrarese}, {Gilbert}, {Kaspi}, {Malkan}, {Maoz}, {Merritt}, {Netzer}, {Onken}, {Pogge}, {Vestergaard}, \& {Wandel}}]{peterson04}
{Peterson}, B.~M., {Ferrarese}, L., {Gilbert}, K.~M., {et~al.} 2004, \apj, 613, 682

\bibitem[{Petrucci {et~al.}(2018)Petrucci, Ursini, Rosa, Bianchi, Cappi, Matt, Dadina, \& Malzac}]{Petrucci_2018}
Petrucci, P.-O., Ursini, F., Rosa, A.~D., {et~al.} 2018, \aap, 611, A59

\bibitem[{{Petrucci} {et~al.}(2013){Petrucci}, {Paltani}, {Malzac}, {Kaastra}, {Cappi}, {Ponti}, {De Marco}, {Kriss}, {Steenbrugge}, {Bianchi}, {Branduardi-Raymont}, {Mehdipour}, {Costantini}, {Dadina}, \& {Lubi{\'n}ski}}]{petrucci13}
{Petrucci}, P.~O., {Paltani}, S., {Malzac}, J., {et~al.} 2013, \aap, 549, A73

\bibitem[{{Porquet, D.} {et~al.}(2018){Porquet, D.}, {Reeves, J. N.}, {Matt, G.}, {Marinucci, A.}, {Nardini, E.}, {Braito, V.}, {Lobban, A.}, {Ballantyne, D. R.}, {Boggs, S. E.}, {Christensen, F. E.}, {Dauser, T.}, {Farrah, D.}, {Garcia, J.}, {Hailey, C. J.}, {Harrison, F.}, {Stern, D.}, {Tortosa, A.}, {Ursini, F.}, \& {Zhang, W. W.}}]{porquet18}
{Porquet, D.}, {Reeves, J. N.}, {Matt, G.}, {et~al.} 2018, A\&A, 609, A42

\bibitem[{Ray {et~al.}(2019)Ray, {STROBE-X Steering Committee \& Working Group}, Brandt, Jaisawal, \& Chenevez}]{ray_2019}
Ray, P., {STROBE-X Steering Committee \& Working Group}, Brandt, S., Jaisawal, G., \& Chenevez, J. 2019, American Astronomical Society. Bulletin (Online), 51

\bibitem[{Reis \& Miller(2013)}]{Reis_2013}
Reis, R.~C., \& Miller, J.~M. 2013, ApJ, 769, L7

\bibitem[{Remillard {et~al.}(2022)Remillard, Loewenstein, Steiner, Prigozhin, LaMarr, Enoto, Gendreau, Arzoumanian, Markwardt, Basak, Stevens, Ray, Altamirano, \& Buisson}]{remillard_2022}
Remillard, R.~A., Loewenstein, M., Steiner, J.~F., {et~al.} 2022, The Astronomical Journal, 163, 130

\bibitem[{{Schlafly} \& {Finkbeiner}(2011)}]{Schlafly11}
{Schlafly}, E.~F., \& {Finkbeiner}, D.~P. 2011, \apj, 737, 103

\bibitem[{{Shakura} \& {Sunyaev}(1973)}]{Shakura1973}
{Shakura}, N.~I., \& {Sunyaev}, R.~A. 1973, \aap, 500, 33

\bibitem[{{Sun} {et~al.}(2018){Sun}, {Grier}, \& {Peterson}}]{Sun_2018}
{Sun}, M., {Grier}, C.~J., \& {Peterson}, B.~M. 2018, {PyCCF: Python Cross Correlation Function for reverberation mapping studies}, Astrophysics Source Code Library, record ascl:1805.032, ascl:1805.032

\bibitem[{Sunyaev \& Tr{\"u}mper(1979)}]{SUNYAEV79}
Sunyaev, R.~A., \& Tr{\"u}mper, J. 1979, Nature, 279, 506–508

\bibitem[{{The Astropy Collaboration} {et~al.}(2013){The Astropy Collaboration}, {Robitaille, Thomas P.}, {Tollerud, Erik J.}, {Greenfield, Perry}, {Droettboom, Michael}, {Bray, Erik}, {Aldcroft, Tom}, {Davis, Matt}, {Ginsburg, Adam}, {Price-Whelan, Adrian M.}, {Kerzendorf, Wolfgang E.}, {Conley, Alexander}, {Crighton, Neil}, {Barbary, Kyle}, {Muna, Demitri}, {Ferguson, Henry}, {Grollier, Fr\'ed\'eric}, {Parikh, Madhura M.}, {Nair, Prasanth H.}, {G\"unther, Hans M.}, {Deil, Christoph}, {Woillez, Julien}, {Conseil, Simon}, {Kramer, Roban}, {Turner, James E. H.}, {Singer, Leo}, {Fox, Ryan}, {Weaver, Benjamin A.}, {Zabalza, Victor}, {Edwards, Zachary I.}, {Azalee Bostroem, K.}, {Burke, D. J.}, {Casey, Andrew R.}, {Crawford, Steven M.}, {Dencheva, Nadia}, {Ely, Justin}, {Jenness, Tim}, {Labrie, Kathleen}, {Lim, Pey Lian}, {Pierfederici, Francesco}, {Pontzen, Andrew}, {Ptak, Andy}, {Refsdal, Brian}, {Servillat, Mathieu}, \& {Streicher, Ole}}]{Astropy_2013}
{The Astropy Collaboration}, {Robitaille, Thomas P.}, {Tollerud, Erik J.}, {et~al.} 2013, A\&A, 558, A33

\bibitem[{{Tie} \& {Kochanek}(2018)}]{tie_2018}
{Tie}, S.~S., \& {Kochanek}, C.~S. 2018, \mnras, 473, 80

\bibitem[{Tortosa {et~al.}(2023)Tortosa, Ricci, Arévalo, Koss, Bauer, Trakhtenbrot, Mushotzky, Temple, Ricci, Rojas Lilayu, Kawamuro, Caglar, Liu, Harrison, Oh, Powell, Stern, \& Urry}]{Tortosa_2023}
Tortosa, A., Ricci, C., Arévalo, P., {et~al.} 2023, \mnras, 526, 1687

\bibitem[{Ursini {et~al.}(2020)Ursini, Dovciak, Zhang, Matt, Petrucci, \& Done}]{ursini20}
Ursini, F., Dovciak, M., Zhang, W., {et~al.} 2020, A\&A, 644, A132

\bibitem[{Uttley {et~al.}(2003)Uttley, Edelson, McHardy, Peterson, \& Markowitz}]{Uttley_2003}
Uttley, P., Edelson, R., McHardy, I.~M., Peterson, B.~M., \& Markowitz, A. 2003, \apj, 584, L53

\bibitem[{{Vaughan} {et~al.}(2003){Vaughan}, {Edelson}, {Warwick}, \& {Uttley}}]{vaughan03}
{Vaughan}, S., {Edelson}, R., {Warwick}, R.~S., \& {Uttley}, P. 2003, \mnras, 345, 1271

\bibitem[{Virtanen {et~al.}(2020)Virtanen, Gommers, Oliphant, Haberland, Reddy, Cournapeau, Burovski, Peterson, Weckesser, Bright, {van der Walt}, Brett, Wilson, Millman, Mayorov, Nelson, Jones, Kern, Larson, Carey, Polat, Feng, Moore, {VanderPlas}, Laxalde, Perktold, Cimrman, Henriksen, Quintero, Harris, Archibald, Ribeiro, Pedregosa, {van Mulbregt}, \& {SciPy 1.0 Contributors}}]{Scipy_2020}
Virtanen, P., Gommers, R., Oliphant, T.~E., {et~al.} 2020, Nature Methods, 17, 261

\bibitem[{Waddell {et~al.}(2023)Waddell, Nandra, Buchner, Wu, Shen, Arcodia, Merloni, Salvato, Dauser, Boller, Liu, Comparat, Wolf, Dwelly, Ricci, Brownstein, \& Brusa}]{waddell_2023}
Waddell, S. G.~H., Nandra, K., Buchner, J., {et~al.} 2023, The eROSITA Final Equatorial Depth Survey (eFEDS): Complex absorption and soft excesses in hard X-ray--selected active galactic nuclei, arXiv:2306.00961

\bibitem[{Wilkins {et~al.}(2016)Wilkins, Cackett, Fabian, \& Reynolds}]{wilkins16}
Wilkins, D.~R., Cackett, E.~M., Fabian, A.~C., \& Reynolds, C.~S. 2016, \mnras, 458, 200

\bibitem[{Wilkins \& Fabian(2013)}]{wilkins13}
Wilkins, D.~R., \& Fabian, A.~C. 2013, \mnras, 430, 247

\bibitem[{{Wilms} {et~al.}(2000){Wilms}, {Allen}, \& {McCray}}]{Wilms_2000}
{Wilms}, J., {Allen}, A., \& {McCray}, R. 2000, \apj, 542, 914

\bibitem[{Yao {et~al.}(2023)Yao, Secunda, Jiang, Greene, \& Villar}]{Yao_2023}
Yao, P.~Z., Secunda, A., Jiang, Y.-F., Greene, J.~E., \& Villar, A. 2023, \apj, 953, 43

\bibitem[{{Zoghbi} {et~al.}(2010){Zoghbi}, {Fabian}, {Uttley}, {Miniutti}, {Gallo}, {Reynolds}, {Miller}, \& {Ponti}}]{zoghbi10}
{Zoghbi}, A., {Fabian}, A.~C., {Uttley}, P., {et~al.} 2010, \mnras, 401, 2419

\bibitem[{Zoghbi \& Miller(2023)}]{Zoghbi_2023}
Zoghbi, A., \& Miller, J.~M. 2023, \apj, 957, 69

\end{thebibliography}
\end{document}